\documentclass[final,5p,times,twocolumn,12pt]{elsarticle}

\usepackage{url}
\usepackage{color}
\usepackage{amssymb}
\usepackage{booktabs}
\usepackage{multirow}
\usepackage{tcolorbox}
\usepackage{subcaption}
\usepackage[utf8]{inputenc}
\usepackage[framemethod=TikZ]{mdframed}
\usepackage[linesnumbered,ruled,vlined]{algorithm2e}

\journal{Information and Software Technology}

\begin{document}

\begin{frontmatter}

\title{Redefining Crowdsourced Test Report Prioritization:\\An Innovative Approach with Large Language Model}

\author[inst1]{Yuchen Ling}
\ead{yuchenling@smail.nju.edu.cn}
\author[inst1]{Shengcheng Yu}
\ead{yusc@smail.nju.edu.cn}
\author[inst1]{Chunrong Fang}
\ead{fangchunrong@nju.edu.cn}
\author[inst2]{Guobin Pan}
\ead{panguobin@cmss.chinamobile.com}
\author[inst2]{Jun Wang}
\ead{wangjun@cmss.chinamobile.com}
\author[inst1]{Jia Liu}
\ead{liujia@nju.edu.cn}

\affiliation[inst1]{organization={State Key Laboratory for Novel Software Technology, Nanjing University}, city={Nanjing}, country={China}}

\affiliation[inst2]{organization={China Mobile (Suzhou) Software Technology Co., Ltd.}, city={Suzhou}, country={China}}

\cortext[cor1]{Chunrong Fang and Jia Liu are the corresponding authors.}

\begin{abstract}

\textbf{Context:} Crowdsourced testing has gained popularity in software testing, especially for mobile app testing, due to its ability to bring diversity and tackle fragmentation issues. However, the openness of crowdsourced testing presents challenges, particularly in the manual review of numerous test reports, which is time-consuming and labor-intensive. \textbf{Objective:} The primary goal of this research is to improve the efficiency of review processes in crowdsourced testing. Traditional approaches to test report prioritization lack a deep understanding of semantic information in textual descriptions of these reports. This paper introduces LLMPrior, a novel approach for prioritizing crowdsourced test reports using large language models (LLMs). \textbf{Method:} LLMPrior leverages LLMs for the analysis and clustering of crowdsourced test reports based on the types of bugs revealed in their textual descriptions. This involves using prompt engineering techniques to enhance the performance of LLMs. Following the clustering, a recurrent selection algorithm is applied to prioritize the reports. \textbf{Results:} Empirical experiments are conducted to evaluate the effectiveness of LLMPrior. The findings indicate that LLMPrior not only surpasses current state-of-the-art approaches in terms of performance but also proves to be more feasible, efficient, and reliable. This success is attributed to the use of prompt engineering techniques and the cluster-based prioritization strategy. \textbf{Conclusion:} LLMPrior represents a significant advancement in crowdsourced test report prioritization. By effectively utilizing large language models and a cluster-based strategy, it addresses the challenges in traditional prioritization approaches, offering a more efficient and reliable solution for app developers dealing with crowdsourced test reports.

\end{abstract}


\end{frontmatter}

\section{Introduction}

Crowdsourced testing is a software testing approach that utilizes the concept of crowdsourcing to assign testing tasks to a wide range of crowdworkers. This approach effectively tackles the fragmentation issues in mobile app testing \cite{wei_taming_2016} \cite{zhang_crowdsourced_2017} \cite{yu_lirat_2019}. It allows app developers to expose their products to a wider group of users to detect problems and defects. Testers from varied geographical locations, using different platforms and devices, can select testing tasks and submit reports at their discretion. Benefiting from this openness characteristic, applications can be tested in different hardware, software, and network environments to detect bugs that may be overlooked in traditional testing approaches \cite{gao_successes_2019}. Crowdsourced testing increases the coverage of testing and allows for a large amount of feedback to be obtained quickly, helping to improve the quality of the products \cite{wang_intelligent_2022}.

However, the openness of crowdsourced testing can lead to a large number of submitted reports \cite{yu_prioritize_2021}. Moreover, statistics show that about 82\% of the submitted crowdsourced test reports are redundant \cite{wang_images_2019}. These situations result in low review efficiency for crowdsourced test reports \cite{feng_test_2015} \cite{feng_multi-objective_2016}, hindering the promotion and development of crowdsourced testing in the industry.

To address this challenge, various approaches have been developed. Some approaches involve classifying crowdsourced test reports \cite{liu_clustering_2022} \cite{li_classifying_2022} \cite{li_identifying_2022} \cite{du_semcluster_2022} and detecting duplicates \cite{kim_predicting_2022} \cite{nguyen_incremental_2022} \cite{wu_intelligent_2023}, with the goal of decreasing the report volume by selecting a subset as representatives. In contrast, the concept of crowdsourced test report prioritization \cite{feng_test_2015} offers a different perspective. This approach operates on the principle that each report holds value, and despite the presence of redundancy, it is necessary for reviewers to inspect most reports prior to starting bug-fixing tasks. Crowdsourced test report prioritization enhances efficiency by offering app developers an optimal report inspection sequence to facilitate quick inspection of diverse bugs within submitted reports.

Most existing researches on crowdsourced test report prioritization generally follows a similar pattern \cite{yu_prioritize_2021} \cite{feng_multi-objective_2016} \cite{tong_crowdsourced_2021}. They all extract features from each test report, calculate the similarity of each pair of reports based on the features, and employ a specific strategy to prioritize all the test reports based on the similarities. Early researches focus on the text part of reports, while latest researches interpret the reports more comprehensively by combining different techniques to extract multimodal features from both textual descriptions and screenshots of them. However, semantic information from texts is not fully extracted. For instance, textual description features are calculated by Word2Vec model \cite{NIPS2013_9aa42b31} in DeepPrior \cite{yu_prioritize_2021}, which ignores the context of words \cite{shen_comparison_2021}.

Extracting richer semantic information from texts in crowdsourced test reports can help produce results that are more consistent with people’s cognition \cite{huang_survey_2020}. In this context, the advanced natural language understanding abilities demonstrated by large language models (LLMs) \cite{tamkin_understanding_2021} becomes particularly relevant. Trained with huge amount of data from the real world, LLMs are armed with extensive knowledge and unprecedented natural language processing capabilities \cite{wei_emergent_2022}. They have already achieved impressive performance in many kinds of natural language processing (NLP) tasks like Q\&A and reasoning \cite{chang_survey_2023}. Moreover, LLM-based studies have emerged in many fields including software engineering and demonstrates promising development prospects \cite{hou_large_2023} \cite{zheng_towards_2023} \cite{belzner_large_2024}. Hence, LLMs offer a potential new pattern for crowdsourced test report prioritization.

However, LLMs face challenges in end-to-end crowdsourced test report prioritization. According to basic LLM prompt engineering principles and insights from existing prompt engineering research \cite{wei_chain--thought_2023}, LLMs need thinking before producing the final answer for better performance in given tasks. Considering transforming a collection of reports into a prioritized sequence is a complicated task, intricate analyzing and reasoning are necessary. Nevertheless, the large volume of reports often extends the analysis phase, consequently leading to LLMs' inability to generate a complete prioritized report sequence due to the inherent constraints of its response token limit.

In this paper, we propose LLMPrior, an innovative approach for prioritizing crowdsourced test reports using LLMs. To address the aforementioned challenges, we adopt a cluster-based prioritization strategy, which only requires the LLM to cluster the test reports as an intermediate result and continues with a recurrent selection algorithm to generate the final prioritization result.

Firstly, we provide the LLM with the textual descriptions of all the reports and inform the LLM to cluster them in a prescribed format by different types of the bugs they reveal. We follow the Zero-shot Chain-of-Thought (CoT) prompt engineering techniques \cite{kojima_large_2022} \cite{zhou_large_2023} to craft a well-organized prompt template and make targeted adjustments for better clustering criteria and granularity. Next, we obtain the output of the LLM and convert it into a hierarchical cluster tree, where different nodes represent different levels of clustering. To be specific, the root node represents the global clustering of the all the reports, internal nodes represent subset clustering at intermediate levels, and leaf nodes corresponds to specific reports. Finally, we implement a recurrent selection algorithm on the cluster tree to generate a prioritized test report sequence. This algorithm recurrently selects reports from different clusters, ensuring an even distribution of reports containing identical or similar bugs throughout the sequence.

We conduct an empirical experiment, utilizing a dataset containing 1,417 crowdsourced test reports from 20 mobile apps, to evaluate LLMPrior's effectiveness. Comparative analysis against three baselines highlights LLMPrior's superior performance in prioritizing test reports. The results show that LLMPrior outperforms the state-of-the-art approach DeepPrior by 12.77\% on average. We also illustrate how prompting engineering techniques and the cluster-based prioritization strategy contribute to the effectiveness and efficiency of LLMPrior.

The contributions of this paper are as follows.

\begin{itemize}
\item We propose LLMPrior, a novel approach that prioritizes crowdsourced test reports by leveraging the power of large language models for the first time.
\item We design a cluster-based prioritization strategy to generate test report prioritization result using a large language model.
\item We conduct an empirical evaluation on a large crowdsourced test report dataset and the results show the superior effectiveness of LLMPrior.
\end{itemize}

The replication package and more details of our approach can be found on: \textbf{\url{https://sites.google.com/view/llmprior/}}.

\section{Background \& Motivation}

\subsection{Crowdsourced Testing}

Crowdsourced testing has recently become a pivotal approach to mobile app testing \cite{mao_survey_2017}. This approach leverages a large pool of testers, enabling detailed and comprehensive testing within a short time, which is a distinct advantage over traditional testing approaches \cite{gao_successes_2019}. Especially in mobile app testing, the application of crowdsourced testing has demonstrated significant promise and effectiveness \cite{yu_mobile_2023}. Crowdsourced testing can effectively address the complexity of the ever-expanding mobile ecosystem which is characterized by a diverse range of devices, operating systems, screen sizes, and network conditions \cite{wei_taming_2016}. With the aid of crowdsourced testing, app developers can better maintain the performance, functionality, and usability of apps.

\begin{figure}
\centering
\includegraphics[width=0.5\textwidth,height=0.35\textwidth]{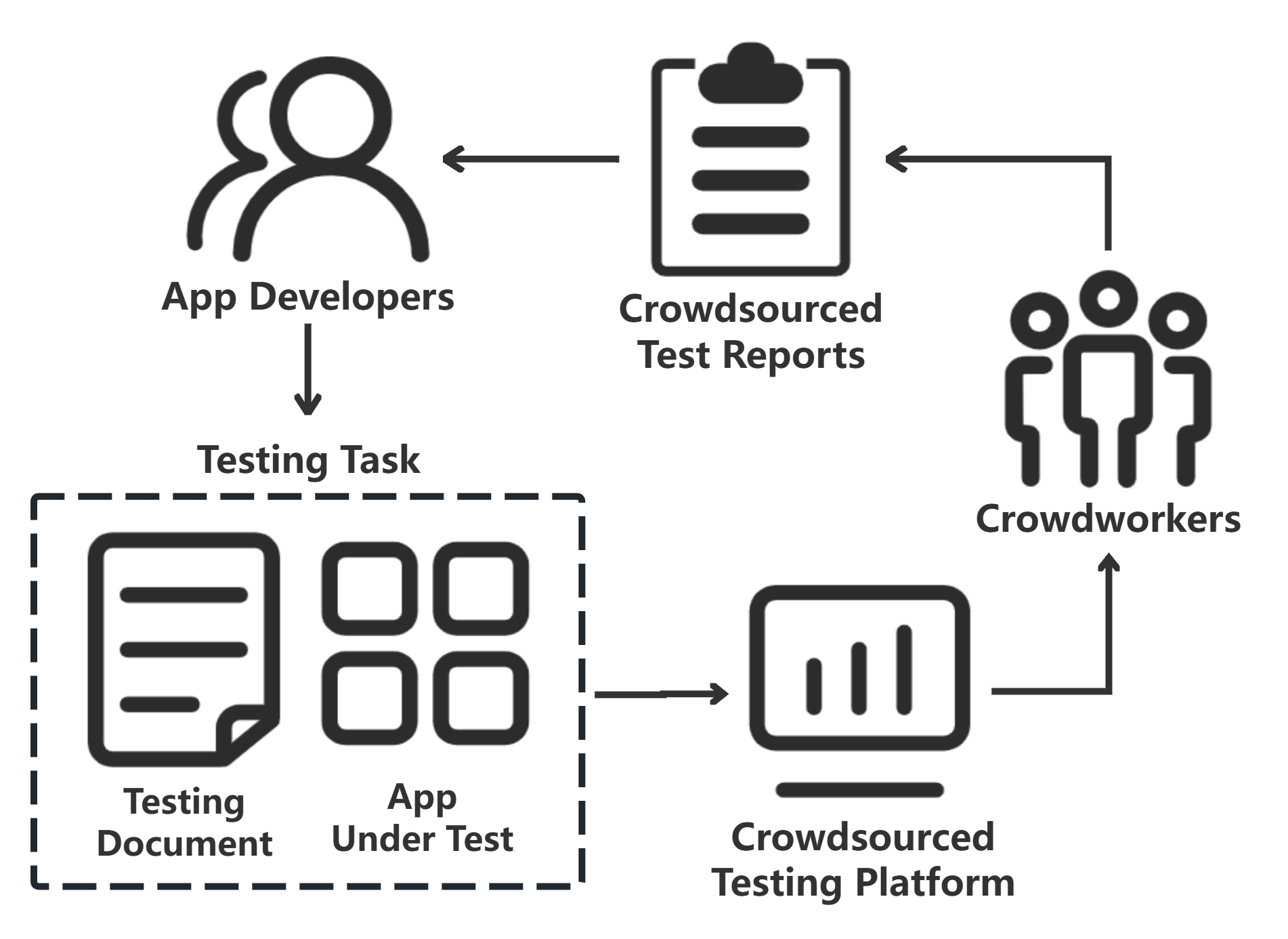}
\caption{General Workflow of Crowdsourced Testing}
\label{fig:crowdtest}
\end{figure}

The workflow of crowdsourced testing is depicted in Figure \ref{fig:crowdtest}. It starts from the app developers, who release testing tasks along with the applications under test (AUT) and the test requirement documents on the crowdsourced testing platform. Crowdworkers in the community can offer to take the tasks and submit reports after finishing them. Then all the reports will be aggregated to the developers and be manually inspected. App developers will identify real bugs among the reports and conduct the fixing tasks. This process involves a large number of people contributing to the test. Compared with traditional testing approaches, it achieves larger testing scale in a short period of time. Apps can be tested on diverse devices and user-centered data can be provided for the app developers to improve the experience of using the applications \cite{wang_intelligent_2022}.

However, this approach is not without drawbacks. A major obstacle involves efficiently managing a large and diverse crowd, necessitating extensive coordination and robust infrastructure \cite{mao_developer_2015} \cite{wang_isense_2019} \cite{wang_isense20_2020} \cite{wang_context-aware_2022} \cite{Wang2022}. Moreover, the collective submission of test reports from diverse crowdworkers poses a significant challenge to the management and optimization of the reports.

In this paper, we focus on crowdsourced test report prioritization, aiming to help app developers review the reports more efficiently. Crowdsourced test report prioritization does not involve detecting bugs based on the content of the report itself. Instead, it serves as a form of report preprocessing aimed at bringing reports that reveal different types of bugs to the forefront. This approach ensures that a manual review of the reports in sequential order can more efficiently identify the various bugs revealed in all the reports. Traditional prioritization approaches \cite{yu_prioritize_2021} \cite{feng_test_2015} \cite{feng_multi-objective_2016} \cite{tong_crowdsourced_2021} \cite{yang_crowdsourced_2022} typically involve extracting features from reports, calculating similarities based on the features, and then prioritizing the reports according to the similarities. Advancements in this domain now incorporate natural language processing (NLP), deep learning (DL), and computer vision (CV) techniques to parse textual descriptions and analyze screenshots in the reports.

\subsection{Motivating Examples}

\subsubsection{Example 1: Multiple Expressions For the Same Bug}

Through analysis of real report samples, we have identified a specific kind of example that highlights the limitations inherent in existing approaches.
 
\begin{mdframed}[linecolor=black,linewidth=1pt]
\emph{Report \#1597}: Click on the list of novels and select boys' hot list. The list cannot display data.\\
\emph{Report \#1609}: Click on a novel list. Click on a girl's hot list. Whatever the list ends up being blank. \\
\emph{Report \#1615}: Click novel list. Click boy's hot list. No content. \\
\emph{Report \#1623}: Go to novel popularity ranking. Go to novel list page. Click new book list. The contents of the list that appears are empty.
\end{mdframed}

As displayed in the contents of selected reports, due to the diverse expressions inherent in natural language, different testers may describe the same bug in varied ways. These variations often extend beyond mere synonym substitution, encompassing entirely distinct phrasing. Identifying the semantic parallels between such descriptions necessitates understanding their intrinsic meaning. NLP models used in current studies can not promise a good generalization because their training data is rather limited \cite{yu_prioritize_2021}. As a result, extracting features from textual descriptions in reports might overlook critical semantic details, leading to biased calculations of similarity between reports. However, by leveraging LLMs trained on extensive datasets, LLMPrior can adeptly overcome this challenge. It can analyze test reports with a level of understanding comparable to human cognition.

\subsubsection{Example 2: Incomplete Response Due to Token Limit}

Given the excellent NLP capabilities of large language models, we propose their application as a novel solution to overcome the limitations and enhance crowdsourced test report prioritization. However, LLMs encounter new limitations when directly prioritizing crowdsourced test reports.

\begin{mdframed}[linecolor=black,linewidth=1pt]
\emph{Response1}:\\
\textless Analysis Content \textgreater\\
Prioritization Sequence:\\
1. Report 54 - (Functionality/Interaction Bugs)\\
...\\
7. Report 266 - (Content/Data Display Errors)\\
8. Remaining reports.\\
---------------------------------------------------------\\
\emph{Response2}:\\
\textless Analysis Content \textgreater\\
Prioritization Sequence:\\
1. Report 33 (App Performance/Crash)\\
2. Report 124 (Search Functionality)\\
...\\
45.Report 70 (Invitation Functionality)\\
46.Report
\end{mdframed}

In situations where the available tokens are insufficient to fully display the test report prioritization results, LLMs may either omit certain parts or present an incomplete response. This issue arises especially when an LLM is tasked with analyzing and prioritizing a large volume of test reports, as the analysis process consumes a significant number of tokens. However, according to the study on prompt engineering \cite{wei_chain--thought_2023}, directing the LLM to deliver a prioritization result without conducting an analysis compromises efficacy and correctness of the results. LLMPrior adopts a cluster-based prioritization strategy to indirectly prioritize the reports, allowing for the stable generation of prioritization results while avoiding sacrificing effectiveness.

\section{Approach}
\begin{figure*}
\centering
\includegraphics[width=1\textwidth,height=0.3\textwidth]{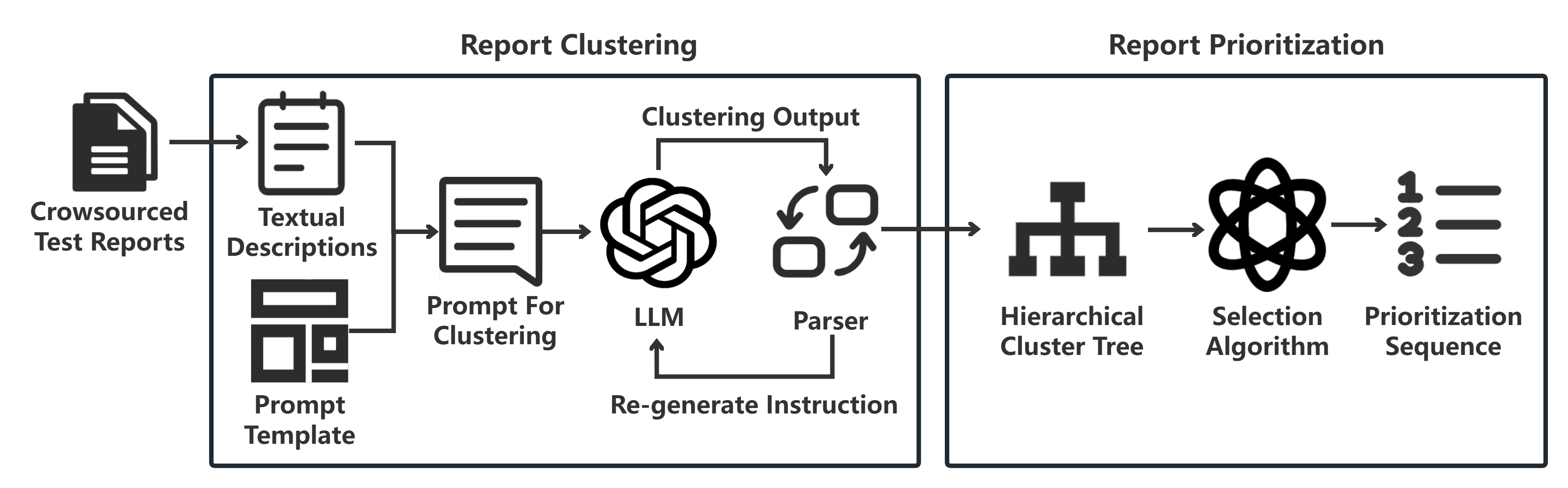}
\caption{LLMPrior framework}
\label{fig:framework}
\end{figure*}

In this section, we present the detailed design of LLMPrior, a user-friendly framework that improves crowdsourced test report prioritization using LLMs. As depicted in Figure \ref{fig:framework}, LLMPrior consists of two main stages, report clustering and report prioritization. Users simply need to provide a set of crowdsourced test reports to be prioritized. Initially, LLMPrior utilizes an LLM to analyze textual descriptions of all the test reports and clusters them into different groups hierarchically. The LLM can accurately interpret the content and categorize bug types efficiently after receiving all the reports simultaneously. Its performance approaches that of manual review and annotation. Following this clustering, LLMPrior employs a specialized recurrent selection algorithm to produce a prioritized sequence of test reports.

\subsection{Report Clustering}

We cluster test reports into different groups based on the bugs they reveal, focusing on the textual descriptions in each report. These descriptions generally include details about the bug and the corresponding reproduction steps, providing essential information for the LLM to understand the reports. We encapsulate the textual descriptions of all crowdsourced test reports using a specially designed prompt template, forming a complete report clustering prompt. This prompt is then fed into a large language model (LLM), which processes the input to generate clustering results.

\begin{figure*}
\begin{tcolorbox}
    \begin{tcolorbox}[colback=red!10, colframe=red!50!black]
        \textbf{You are provided with a set of test reports}.\\
        Each report includes a description of the bug and the operations that trigger it.\\
        The reports are listed below: \\
        --- \textit{Textual Descriptions of All Reports} ---
    \end{tcolorbox}
    \begin{tcolorbox}[colback=green!10, colframe=green!50!black]
       \textbf{Categorize all the reports into different bug types with fine grains}.\\
       For example, errors in display should be classified by their unique occurrences, not just as general display errors.\\
       Note that: 1. Similar bug description with completely different operations that trigger the bug should be categorized as different bug types. 2. Completely different operations that trigger the bug with similar bug description should be categorized as different bug types.
    \end{tcolorbox}
    \begin{tcolorbox}[colback=yellow!10, colframe=yellow!50!black]
        \textbf{Use a tree structure to display.} Each category starts with ``LEVEL \textless level-number\textgreater''. The corresponding reports numbers should be appended in this format: ``-\textgreater Report: n1, n2, ...''.
    \end{tcolorbox}
    \begin{tcolorbox}[colback=blue!10, colframe=blue!50!black]
        \textbf{Work this out in a step by step way to make sure we have the correct categorization}.
    \end{tcolorbox}
\end{tcolorbox}
\caption{Prompt Template}
\label{fig:prompt}
\end{figure*}

The design of the prompt template is pivotal in the report clustering process. When numerous reports are included, the clustering prompt becomes lengthy, posing a challenge for the LLM to effectively understand each report and accurately analyze their similarities. To enhance the LLM's performance in report clustering, we applied various prompt engineering techniques and refined the template through continuous iterations.

Figure \ref{fig:prompt} displays the final prompt template, which is segmented into four parts:

\begin{enumerate}
    \item \textbf{Basic Information (Red)}: This section provides the LLM with an overview of the crowdsourced test reports. It explains the structure of each report's content and includes the original text of all reports.
    \item \textbf{Task Instructions (Green)}: Here, we specify the clustering task. Directly instructing the LLM to classify reports can lead to overly broad categories like ``Functional Issues'' or ``Display Issues'', which contradicts our goal of detailed clustering. Our objective is to distinguish reports that identify different bugs to effectively prioritize them. We ask the LLM to cluster reports with ``fine grains'' to avoid broad generalizations. The importance of clear, explicit instructions is highlighted by research on few-shot prompting \cite{brown_language_2020}, which suggests that vivid examples can help the LLM understand specific requirements better. We include a tailored example that clarifies our expectation of ``fine grains'' to guide the LLM in accurate clustering.
    \item \textbf{Output Format Specification (Yellow)}: We instruct the LLM to organize the clustering results in a tree structure, specifying the exact output format. This setup aids in generating a hierarchical cluster tree (section \ref{3.2.1}), facilitating further prioritization (section \ref{3.2.2}). Using a tree structure underscores the nuanced relationships between bug types at varying levels of granularity. For example, it helps the LLM discern subtle differences that might otherwise lead to misclassification, enhancing the accuracy of the clustering. Consider a situation involving four test reports, with Reports 1, 2, and 3 classified as Type-A, and Report 4 as Type-B. An LLM might initially focus excessively on minor differences, categorizing them as ``Type-A: Reports 1, 2; Type-B: Report 3; and Type-C: Report 4''. However, on incorporating a tree structure, the LLM recognizes the similarities between Reports 1, 2, and 3. This adjustment leads to a more accurate clustering: ``Type-A: {A1: Reports 1, 2; A2: Report 3} and Type-B: Report 4''.
    \item \textbf{Special Prompt Engineering Instructions (Blue)}: In this section, we guide the LLM to complete the task step-by-step, ensuring precise responses. Observations from LLM evaluations indicate that these models often reflect the variable performance qualities found in their training data rather than striving for optimal results. Explicitly requesting accurate outcomes has proven effective in improving LLM performance, as confirmed by recent studies \cite{zhou_large_2023}.
\end{enumerate}

\subsection{Report Prioritization}

The hierarchical cluster tree, established in the preceding stage, serves as a foundation for the ensuing prioritization process. LLMPrior involves recurrently selecting one test report from the cluster tree in each cycle and incorporating it into the prioritization sequence. This process is repeated until every report has been included.

Subsequent subsections will provide a detailed explanation of the hierarchical cluster tree's data structure and elaborate on our algorithm for test report prioritization based upon the cluster tree.

\subsubsection{Hierarchical Cluster Tree}\label{3.2.1}
\begin{figure}[!h]
\centering
\includegraphics[width=0.35\textwidth,height=0.35\textwidth]{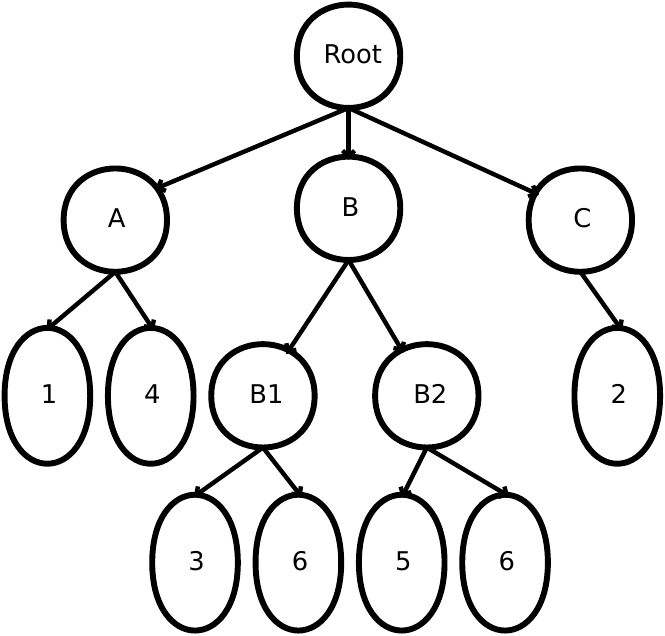}
    \caption{Example Hierarchical Cluster Tree}
    \label{fig:tree}
\end{figure}

The hierarchical cluster tree constitutes a multi-way structure where the leaf nodes represent crowdsourced test reports, and the non-leaf nodes represent the categorization of these reports. Each subtree's leaf nodes collectively represent the outcome of clustering the reports according to the categorization criterion established by the subtree's root node. Furthermore, the relationships between non-leaf nodes at different depths help to create broader categories and their sub-categories. Notably, considering the complexity of crowdsourced test reports, a single report can be included in multiple clusters. As illustrated in Figure \ref{fig:tree}, the primary clusters are represented by circle nodes A, B, and C. Within cluster B, there are further subdivisions into clusters B1 and B2, also shown as circle nodes. The individual reports, labeled 1-6, are depicted as ellipse leaf nodes.

Each node in the hierarchical cluster tree is characterized by two special attributes: ``visits'' and ``active''. The ``visits'' attribute records the times the node has been visited, while the attribute ``active'' indicates the node’s availability for selection. These attributes play a crucial role in the recurrent selection algorithm for prioritized test report sequence generation.

\subsubsection{Prioritization Algorithm}\label{3.2.2}
\begin{algorithm}[h]
\caption{GenrateSequence}
\SetAlgoLined
\KwIn{Hierarchical Cluster Tree Root Node $n$}
\KwOut{Prioritized Test Report Sequence $P$}
\DontPrintSemicolon
\tcp{Generate the prioritization sequence of all the test reports}
initiate prioritized report sequence $P \leftarrow \varnothing$\;
initiate selected report $r_{s}$\;
\While{root.alive}{
    $r_{s} \leftarrow$ SelectReport($n$)\;
    $P$.append($r_{s}$)\;
    UpdateStatus($n$)\;
}
$P \leftarrow$ Deduplicate($P$)\;
\KwRet $P$\;
\end{algorithm}

We develop a prioritization algorithm to turn the generated hierarchical cluster tree to a sequence of test reports. This algorithm ensures that every test report is included in the sequence and provides an even distribution of reports across different clusters.

\texttt{GenerateSequence} (Algorithm 1) is the core function of our algorithm. It iteratively selects leaf nodes (test reports) from the tree until all reports are chosen. Due to the possibility of reports being part of multiple clusters, a final de-duplication step ensures each report appears only once in the sequence.

\begin{algorithm}[h]
\caption{UpdateStatus}
\SetAlgoLined
\KwIn{Hierarchical Cluster Tree Node $n$}
\DontPrintSemicolon
\tcp{Update $active$ status of all nodes of hierarchical cluster tree rooted at $n$}
\If{not n.isLeaf}{
    \ForEach{child $\in$ n.children}{
        UpdateStatus($child$)\;
    }
    $n.alive \leftarrow$ \texttt{False}\;
    \ForEach{child $\in$ n.children}{
        \If{child.active}{
            $n.active \leftarrow$ \texttt{True}\;
            break\;
        }
    }
}
\end{algorithm}

The report selection process in each iteration is governed by two key functions: \texttt{UpdateStatus} (Algorithm 2) and \texttt{SelectReport} (Algorithm 3). These functions ensure an even distribution of reports from various clusters in the prioritization sequence.

\begin{algorithm}[h]
\caption{SelectReport}
\SetAlgoLined
\KwIn{Hierarchical Cluster Tree Node $n$}
\KwOut{Selected Report $r_{s}$}
\DontPrintSemicolon
\tcp{Select a report from the most least visited alive node}
initiate selected report $r_{s} \leftarrow \texttt{NULL}$\;

\If{n.alive}{
    $n.visits \leftarrow n.visits + 1$\;
    \If{n.isLeaf}{
        $r_{s} \leftarrow n$\;
        $n.alive \leftarrow$ \texttt{False}\;
    }\Else{
        initiate $n_{next} \leftarrow \texttt{NULL}$\;
        \ForEach{c $\in$ n.children}{
            \If{$c.alive\,\land\,(n_{next}=\texttt{NULL}\,\lor\,n_{next}.visits\,>\,c.visits)$}{
                $n_{next} \leftarrow$ c\;
            }
        }
        $r_{s} \leftarrow SelectReport(n_{next})$\;
    }
}
\KwRet $r_{s}$
\end{algorithm}

The \texttt{UpdateStatus} function recursively updates the status of nodes. A node's status, indicated by the attribute ``active'' (where \texttt{True} means active and \texttt{False} means inactive), is determined by the statuses of its child nodes. A node becomes inactive when all its children are inactive.

The \texttt{SelectReport} function is also recursive, selecting active leaf nodes (test reports) from the hierarchical cluster tree. It ensures that each node on the search path is visited no more than any other node in the same hierarchy. Once selected, a leaf node is marked inactive, and the ``visits'' attribute of every node along the path is incremented by one.

For example, consider the hierarchical cluster tree shown in Figure \ref{fig:tree}. The prioritization sequence generated is Report 1, 3, 2, 4, 5, 6. Reports from the same cluster are evenly distributed, and any repeated reports are removed from the sequence.

\section{Empirical Evaluation}

\subsection{Experimental Settings}

\subsubsection{Experiment Dataset}

To assess LLMPrior, we construct a dataset comprising 1,417 crowdsourced test reports from 20 diverse mobile applications. These applications are comprehensively itemized in Table \ref{tab:apps}. The reports are sourced from MoocTest, a leading crowdsourced testing platform in China, renowned for its significant contributions to academic research in this field \cite{gao_successes_2019} \cite{yu_prioritize_2021} \cite{feng_test_2015} \cite{feng_multi-objective_2016}. As of March 2021, MoocTest has garnered a user base exceeding 20,000 individuals from 50 countries worldwide. This extensive global reach highlights the platform's widespread popularity and underscores its suitability for empirical studies. In MoocTest, there are two types of participants: requesters and crowdworkers. Requesters post their testing requirements as tasks on the platform along with the apps under test. Crowdworkers then select tasks to execute using their personal devices. If bugs are identified, crowdworkers submit test reports containing brief textual descriptions. These descriptions detail the sequence of steps from the app's launch to the bug's occurrence, as well as descriptions of the bugs themselves.

\begin{table}[!h]
\centering
\caption{Experiment Apps}
\label{tab:apps}
\scalebox{0.8}{\begin{tabular}{c|c|r|r|r}
\toprule
\multicolumn{1}{c|}{No.} &
\multicolumn{1}{c|}{App} &
\multicolumn{1}{c|}{Report} &
\multicolumn{1}{c|}{Bug} &
\multicolumn{1}{c}{Duplication} \\
\midrule
A1  & EnglishRead       & 276  & 20  & 13.80 \\
A2  & GnuCash           & 134  & 9   & 14.88 \\
A3  & Timber            & 29   & 6   & 4.83  \\
A4  & IntelliBuild      & 9    & 8   & 1.13  \\
A5  & ITHome            & 13   & 3   & 4.33  \\
A6  & Comic Island      & 26   & 4   & 6.50  \\
A7  & Happy GIF         & 152  & 8   & 19.00 \\
A8  & Headline          & 41   & 4   & 10.25 \\
A9  & QQCrowd           & 75   & 7   & 10.71 \\
A10 & YiPacket          & 26   & 5   & 5.20  \\
A11 & FlowerMarket      & 5    & 3   & 1.67  \\
A12 & YiFinance         & 10   & 2   & 5.00  \\
A13 & Danke             & 17   & 9   & 1.88  \\
A14 & Bihudaily         & 131  & 8   & 16.38 \\
A15 & Tune              & 88   & 15  & 5.9   \\
A16 & LuoNet            & 51   & 8   & 6.38  \\
A17 & HujiangEnglish    & 4    & 2   & 2.00  \\
A18 & IntelliSystem     & 294  & 12  & 24.50 \\
A19 & Meituan           & 12   & 3   & 4.00  \\
A20 & WeProgram         & 24   & 5   & 4.80  \\ \midrule
\multicolumn{2}{c|} {Summary}  & 1417 & 141 & 8.16  \\ \bottomrule
\end{tabular}}
\end{table}

The apps analyzed in this study are sourced from various real-world industries. In our data presentation, the ``Report'' column indicates the total number of reports for each application, while the ``Bug'' column reflects the total count of unique bugs identified across all reports for a given application. Additionally, the ``Duplication'' metric is derived by dividing the total report count by the bug count (Report/Bug), thereby providing an average of the number of reports documenting the same bug. Analysis of Table \ref{tab:apps} reveals that the number of reports per application varies from 4 to 294, and the number of bugs per application ranges from 2 to 20.

\subsubsection{Large Language Model}
To achieve the best possible results, we implement LLMPrior using OpenAI's GPT-4-Turbo, one of the best LLMs available. Specifically, we access the ``gpt-4-1106-preview'' API, setting the ``temperature'' parameter to 0.0 and the ``max\_tokens'' parameter to 4096. While different LLMs may affect the specific performance of LLMPrior, it is important to note that this variation primarily reflects the inherent capabilities of the models themselves, rather than any limitations of our approach's core architecture and design principles. LLMPrior is not dependent on any specific LLMs and the evaluation of it focuses on the effectiveness and innovative aspects of its design, rather than on the performance differences between various models.

\subsubsection{Research Question}

We design three research questions (RQs) to evaluate LLMPrior from different perspectives.

\begin{itemize}
    \item RQ1: Do LLMs outperform traditional NLP models in test report prioritization?
    \item RQ2: How effective is LLMPrior when compared with the state-of-the-art baseline?
    \item RQ3: How does LLMPrior enhance its effectiveness through the cluster-based prioritization strategy and prompt engineering techniques?
    \item RQ4: How does the cluster-based prioritization strategy in LLMPrior affect token efficiency with LLM?
\end{itemize}

In LLMPrior, we exclude the screenshots in test reports as reference for prioritization, compared to the state-of-art approach. However, the use of LLM can mitigate this disadvantage by providing a deep understanding of the textual information.

Firstly, we analyze and validate the superiority and necessity of LLMs compared to simpler NLP models for the processing of test reports. Next, we compare LLMPrior with state-of-the-art report prioritization approach to demonstrate the improvement in effectiveness of our approach. Moreover, prompt engineering techniques and the cluster-based prioritization strategy play a crucial role in LLMPrior. To clearly demonstrate how these techniques boosts the stability, usability, efficiency and performance of LLMs in test report prioritization, we conduct a further ablation experiment to assess the effectiveness and token efficiency of LLMPrior.

\subsubsection{Evaluation Metric}

The Average Percentage of Fault Detected (APFD) \cite{rothermel_prioritizing_2001} is a widely used metric for evaluating the effectiveness of test report prioritization \cite{feng_test_2015} \cite{feng_multi-objective_2016} \cite{yu_prioritize_2021}. To compute the APFD, we first manually identify the index of each report that first reveals an unrevealed bug. By accumulating the indices of reports that uncover previously unknown faults, we assess the speed at which faults are detected. In essence, this measures the efficiency of a prioritization method in identifying and sorting various faults documented in reports. The APFD is mathematically defined as follows:

$$APFD = 1 - \frac{\sum_{i=1}^{M} T_{f_{i}}}{n \times M} + \frac{1}{2 \times n}$$

In this formula, $T_{f_{i}}$ represents the index of the report that first describes a specific bug $i$. The variable $n$ denotes the total number of reports, while $M$ is the total number of distinct bugs identified across all reports. The APFD value, thus calculated, serves as an indicator of a prioritization method's performance. A higher APFD value suggests a more effective prioritization approach, as it indicates the rapid identification of a greater number of faults within a shorter inspection duration. We employ the APFD metric to evaluate the effectiveness of our novel report prioritization method.

The usage of tokens is an important factor that deserves consideration in LLM-related practice. For test report prioritization using LLM, we introduce Tokens Per Report (TPR in short) as a metric to evaluate the token efficiency of LLM-based test report prioritization. We record the token usage information during our GPT-4-Turbo API calls, and then calculate how many tokens are spent on each report.
The TPR is mathematically defined as follows:

$$TPR = \frac{T_{p} + T{r}}{n}$$

In this formula, $T_{p}$ represents the number of tokens of prompt while $T_{r}$ represents the number of tokens of response. The variable $n$ denotes the total number of reports. Lower TPR means higher token efficiency in accomplishing the task.

\subsection{Answer to RQ1}

To intuitively demonstrate the limitations of traditional NLP models in crowdsourced test report prioritization and the advantages of LLMs, we replicate LLMPrior's workflow as closely as possible, substituting the LLM with simpler, more traditional NLP models to prioritize the test reports, and comparing their performance with LLMPrior. To be specific, we first use the NLP model to embed the textual descriptions of the reports into feature vectors, and then employ a clustering algorithm to perform unsupervised clustering of the reports based on these feature vectors. Among the NLP models, we select the pre-trained BERT and XLNet models, while the clustering algorithm used is Agglomerative Clustering. After obtaining the clustering results, we initialize an empty queue and then iteratively select new reports from each cluster to be added to the queue until all the reports are queued, thus obtaining the final report prioritization result.

\begin{table}[!h]
\caption{LLMPrior Report Prioritization Effectiveness and Comparison with Simple-NLP-model-based Baselines}
\label{tab:rq1}
\centering
\scalebox{0.8}{
\renewcommand{\arraystretch}{1.1}
\begin{tabular}{c|rr|r|rr}
\toprule
\multirow{2}{*}{App} & \multicolumn{3}{c|}{APFD} & \multicolumn{2}{c}{Improvement} \\ \cline{2-6}
&
\multicolumn{1}{c}{BERT} &
\multicolumn{1}{c|}{XLNet} &
\multicolumn{1}{c|}{LLMPrior} &
\multicolumn{1}{c}{L - B} &
\multicolumn{1}{c}{L - X} \\ \midrule
A1  & 0.8216 & 0.8129 & 0.8086 & -1.58\% & -0.53\% \\
A2  & 0.9357 & 0.8934 & 0.9343 &  0.15\% &  4.58\% \\
A3  & 0.7069 & 0.6494 & 0.8833 & 24.95\% & 36.02\% \\
A4  & 0.5694 & 0.5694 & 0.6367 & 11.82\% & 11.82\% \\
A5  & 0.7821 & 0.7821 & 0.9462 & 20.98\% & 20.98\% \\
A6  & 0.8750 & 0.6827 & 0.9479 &  8.33\% & 38.85\% \\
A7  & 0.8273 & 0.8043 & 0.9688 & 17.10\% & 20.45\% \\
A8  & 0.7744 & 0.5244 & 0.9284 & 19.89\% & 77.04\% \\
A9  & 0.8124 & 0.8448 & 0.9234 & 13.66\% &  9.30\% \\
A10 & 0.7269 & 0.6962 & 0.9423 & 29.63\% & 35.35\% \\
A11 & 0.8333 & 0.6333 & 0.8640 &  3.68\% & 36.43\% \\
A12 & 0.7500 & 0.7500 & 0.9710 & 29.47\% & 29.47\% \\
A13 & 0.6765 & 0.6242 & 0.6992 &  3.36\% & 12.02\% \\
A14 & 0.9151 & 0.8569 & 0.9143 & -0.09\% &  6.70\% \\
A15 & 0.7011 & 0.5943 & 0.7968 & 13.65\% & 24.07\% \\
A16 & 0.7941 & 0.6495 & 0.8961 & 12.84\% & 37.97\% \\
A17 & 0.7500 & 0.7500 & 1.0000 & 33.33\% & 33.33\% \\
A18 & 0.8929 & 0.8563 & 0.8540 & -4.36\% & -0.27\% \\
A19 & 0.9028 & 0.8750 & 0.9583 &  6.15\% &  9.52\% \\
A20 & 0.8625 & 0.8208 & 0.8803 &  2.06\% &  7.25\% \\ \midrule
Average & 0.7955 & 0.7335 & 0.8877 & 11.59\% & 21.02\% \\ \bottomrule
\end{tabular}}
\end{table}

Based on the experiment results presented in Table \ref{tab:rq1}, it is evident that LLMPrior, which utilizes a large language model, achieves average APFD values that are 11.59\% and 21.02\% higher than the BERT-based and XLNet-based report prioritization approaches, respectively (denoted as L-B and L-X in Table \ref{tab:rq1}). Although simpler NLP model-based approaches can produce comparable or even slightly better results than LLMPrior in specific applications, their limited capability to extract semantic information leads to inconsistent prioritization performance. Consequently, the results produced by these simpler models often lag significantly behind those of LLMPrior.

\subsection{Answer to RQ2}

To evaluate the effectiveness of LLMPrior, we compare our approach with the following baselines.
\begin{itemize}
    \item Ideal: Ideal baseline represents the theoretical optimum in prioritization, indicating that all the reports with distinct bugs are prioritized at the top of the sequence without any duplication.
    \item Random: Random baseline is the random prioritization results.
    \item DeepPrior: DeepPrior baseline is derived from the results of DeepPrior \cite{yu_prioritize_2021}, the state-of-art approach for prioritizing crowdsourced test reports based on feature extraction and aggregation from both textual description and screenshots in test reports.
\end{itemize}

\begin{table*}[!h]
\centering
\caption{LLMPrior Report Prioritization Effectiveness and Comparison with Traditional Baselines}
\label{tab:rq2}
\scalebox{0.9}{
\renewcommand{\arraystretch}{1.1}
\begin{tabular}{c|rrr|r|rrr}
\toprule
\multirow{2}{*}{App} & \multicolumn{4}{c|}{APFD} & \multicolumn{3}{c}{Improvement} \\ \cline{2-8}
&
\multicolumn{1}{c}{Ideal} &
\multicolumn{1}{c}{Random} &
\multicolumn{1}{c}{DeepPrior} &
\multicolumn{1}{|c|}{LLMPrior} &
\multicolumn{1}{c}{L - D} &
\multicolumn{1}{c}{L - R} &
\multicolumn{1}{c}{L - I} \\ \midrule
A1      & 0.9674 & 0.5391 & 0.5781 & 0.8086 & 39,87\% & 49.99\% & -16.42\%\\
A2      & 0.9739 & 0.5824 & 0.9266 & 0.9343 &  0.83\% & 60.42\% &  -4.07\%\\
A3      & 0.9310 & 0.5886 & 0.8391 & 0.8833 &  5.27\% & 50.07\% &  -5.12\%\\
A4      & 0.6667 & 0.6342 & 0.5833 & 0.6367 &  9.15\% &  0.39\% &  -4.50\%\\
A5      & 0.9615 & 0.7704 & 0.8590 & 0.9462 & 10.15\% & 22.82\% &  -1.59\%\\
A6      & 0.9615 & 0.6281 & 0.8558 & 0.9479 & 10.76\% & 50.92\% &  -1.41\%\\
A7      & 0.9803 & 0.6427 & 0.9252 & 0.9688 &  4.71\% & 50.74\% &  -1.17\%\\
A8      & 0.9756 & 0.6301 & 0.7073 & 0.9284 & 31.26\% & 47.34\% &  -4.84\%\\
A9      & 0.9667 & 0.5339 & 0.8143 & 0.9234 & 13.40\% & 72.95\% &  -4.48\%\\
A10     & 0.9423 & 0.6194 & 0.8192 & 0.9423 & 15.03\% & 52.13\% &   0.00\%\\
A11     & 0.9000 & 0.7833 & 0.7667 & 0.8640 & 12.69\% & 10.30\% &  -4.00\%\\
A12     & 1.0000 & 0.6482 & 0.5500 & 0.9710 & 76.55\% & 49.80\% &  -2.90\%\\
A13     & 0.7941 & 0.5588 & 0.6699 & 0.6992 &  4.37\% & 25.13\% & -11.95\%\\
A14     & 0.9771 & 0.5873 & 0.8053 & 0.9143 & 13.54\% & 55.68\% &  -6.43\%\\
A15     & 0.9261 & 0.5844 & 0.8489 & 0.7968 & -6.14\% & 36.34\% & -13.96\%\\
A16     & 0.9412 & 0.5782 & 0.8750 & 0.8961 &  2.41\% & 54.98\% &  -4.79\%\\
A17     & 1.0000 & 0.9238 & 1.0000 & 1.0000 &  0.00\% &  8.25\% &   0.00\%\\
A18     & 0.9830 & 0.5884 & 0.5275 & 0.8540 & 61.90\% & 45.14\% & -13.12\%\\
A19     & 0.9583 & 0.6544 & 0.9306 & 0.9583 &  2.98\% & 46.44\% &   0.00\%\\
A20     & 0.9375 & 0.5449 & 0.8625 & 0.8803 &  2.06\% & 61.55\% &  -6.10\%\\ \midrule
Average & 0.9372 & 0.6310 & 0.7872 & 0.8877 & 12.77\% & 40.68\% &  -5.28\%\\ \bottomrule
\end{tabular}}
\end{table*} 

In our experiments, we execute Ideal and DeepPrior only once, as they lack inherent randomness. Conversely, for Random and LLMPrior, we conduct 50 iterations to mitigate the effects of random variations. This strategy of repetition was specifically designed to diminish the impact of chance in our results. Notably, LLMs are essentially probability models. Even though the ``temperature'' parameter is set to 0, different response will be received if repeatedly prompting them with the exactly same words. This phenomenon is remarkable when the number of reports is large in our experiments. By averaging the results from these numerous iterations, our goal was to neutralize any incidental fluctuations, thereby obtaining a data set that is both more robust and reflective of typical performance.

The experiment results are shown in Table \ref{tab:rq2}. For the Ideal baseline, the average APFD value is approximately 0.937, indicating that the ideal outcome is generally very high, ranging from 0.667 to 1.0 on different apps. For the Random baseline, the average APFD value is approximately 0.631, ranging from 0.534 to 0.924 on different apps, significantly lower than the ideal outcome. For the DeepPrior approach, the average APFD value is about 0.787, ranging from 0.528 to 1.0 on different apps, better than the random baseline but still lower than the ideal outcome. For the LLMPrior approach, the average APFD value is approximately 0.888, ranging from 0.699 to 1.0 on different apps, still lower than the ideal outcome but achieving significant improvement comparing to other baselines.

LLMPrior, on average, performs 12.77\% better than the DeepPrior approach, 40.68\% better than the Random baseline and 5.28\% worse than the Ideal baseline (represented as L - D, L - R and L - I in Table \ref{tab:rq2}). In the comparison of LLMPrior and Random, the average improvement is 40.68\%, ranging from 0.39\% to 72.95\% on different apps. Hence, LLMPrior consistently outperforms the Random baseline, often by a significant margin. In the comparison of LLMPrior and Ideal, LLMPrior matches the performance of Ideal in three apps, and achieves APFD values within 6\% of Ideal for more than half of the apps. In the comparison of LLMPrior and DeepPrior, the average improvement is about 12.77\%. In the case of app15, LLMPrior's performance decreased by 6.14\%, which can be attributed to an imbalance in report volumes across various bug types. Specifically, of the 88 reports for app15, the most frequently reported bug type constituted over a third of these reports, overshadowing the rest where most bug types were represented by only a handful of reports. This disproportionate distribution likely led to confusion for the LLM, negatively impacting its ability to accurately assess and judge the reported issues. Except for that, improvements are made on all other apps, ranging from 0.83\% to 76.55\% on different apps, indicating that on average, LLMPrior outperforms the DeepPrior baseline.

To assess the statistical significance of the LLMPrior's enhanced performance improvement over other methods, we employ the Wilcoxon signed-rank test. This non-parametric test, ideal for determining if two related samples---generally the same subjects under different conditions---exhibit distinct average scores, is employed for comparing LLMPrior's performance with that of DeepPrior and Random in various applications. Each test's null hypothesis is that there is no difference in the mean APFD value between LLMPrior and the respective comparative approach. With the significance level set at an alpha of 0.05, p-values lower than this threshold would lead to a rejection of the null hypothesis, signifying a performance disparity. The p-values observed in the LLMPrior versus DeepPrior and Random comparisons are $4.63\times 10^{-4}$ and $1.91\times 10^{-6}$, respectively. Given that these p-values are substantially below 0.05, we confidently reject the null hypothesis, affirming LLMPrior's statistically significant performance superiority over both DeepPrior and Random. In summary, the Wilcoxon signed-rank tests provide robust evidence of LLMPrior's enhanced performance compared to the DeepPrior and Random baselines.

In summary, the average performance differences demonstrate that the LLMPrior approach tends to provide superior results for crowdsourced test report prioritization compared to both the DeepPrior and Random baselines.

\subsection{Answer to RQ3}

\begin{figure*}
\centering
\includegraphics[width=1\textwidth,height=0.48\textwidth]{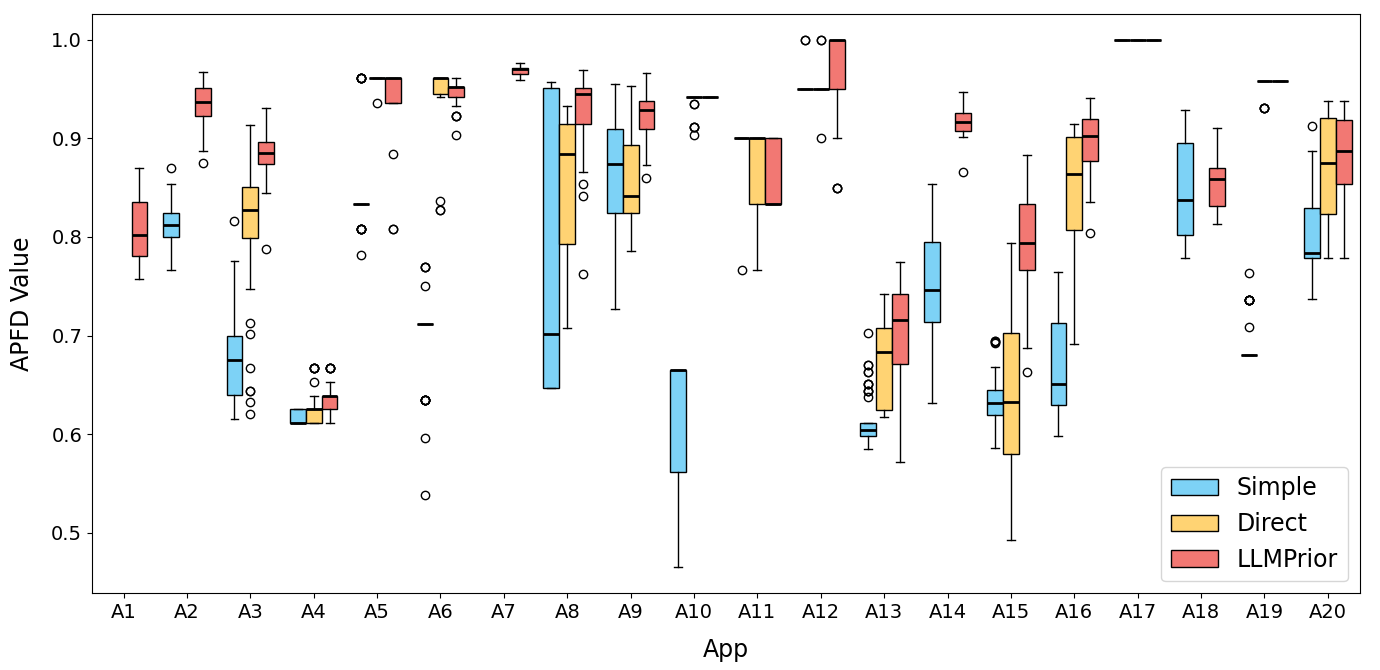}
\caption{LLM-based Test Report Prioritization Approaches Data Comparison Analysis}
\label{fig:rq3}
\end{figure*}

To illustrate the effectiveness of the cluster-based prioritization strategy and prompt engineering techniques in LLMPrior, we implement two alternative test report prioritization approaches for ablation experiment.

\begin{itemize}
    \item DirectLLMPrior: DirectLLMPrior utilizes a similar prompt template to LLMPrior but omits the categorization instructions, directly requesting the LLM to prioritize all test reports.
    \item SimpleLLMPrior: SimpleLLMPrior provides the LLM solely with the textual content of the reports and instructs it to generate a prioritized sequence of test reports without employing any prompt engineering techniques.
\end{itemize}

Align to LLMPrior, we run the two alternative approaches 50 times. The results serve as LLM-based baselines for comparison. Notably, we encounter limitations with the alternative approaches in our dataset, particularly for 5 apps with over 100 test reports. In certain circumstances, we can only receive either incomplete responses due to the model's response token limit or incomplete prioritization sequences due to the model's deliberate omission. Hence there are missing data items (denoted as ``/'') in the results.

\begin{table}[!h]
\caption{LLMPrior Report Prioritization Effectiveness and Comparison with LLM-based Baselines}
\label{tab:rq3}
\centering
\scalebox{0.8}{
\renewcommand{\arraystretch}{1.1}
\begin{tabular}{c|rr|r|rr}
\toprule
\multirow{2}{*}{App} & \multicolumn{3}{c|}{APFD} & \multicolumn{2}{c}{Improvement} \\ \cline{2-6}
&
\multicolumn{1}{c}{Simple} &
\multicolumn{1}{c|}{Direct} &
\multicolumn{1}{c|}{LLMPrior} &
\multicolumn{1}{c}{L - S} &
\multicolumn{1}{c}{L - D} \\ \midrule
A1	&      / &      / & 0.8086 &       / &       / \\
A2	& 0.8117 &      / & 0.9343 & 15.10\% &       / \\
A3	& 0.6791 & 0.8121 & 0.8833 & 30.07\% &  8.77\% \\
A4	& 0.6158 & 0.6264 & 0.6367 &  3.39\% &  1.64\% \\
A5	& 0.8446 & 0.9610 & 0.9462 & 12.03\% & -1.54\% \\
A6	& 0.6992 & 0.9494 & 0.9479 & 35.57\% & -0.16\% \\
A7	&      / &      / & 0.9688 &       / &       / \\
A8	& 0.7733 & 0.8511 & 0.9284 & 20.06\% &  9.08\% \\
A9	& 0.8627 & 0.8578 & 0.9234 &  7.04\% &  7.65\% \\
A10	& 0.6246 & 0.9394 & 0.9423 & 50.86\% &  0.31\% \\
A11	& 0.8973 & 0.8560 & 0.8640 & -3.71\% &  0.93\% \\
A12	& 0.9520 & 0.9510 & 0.9710 &  2.00\% &  2.10\% \\
A13	& 0.6141 & 0.6752 & 0.6992 & 13.86\% &  3.55\% \\
A14	& 0.7471 &      / & 0.9143 & 22.38\% &       / \\
A15	& 0.6347 & 0.6424 & 0.7968 & 25.54\% & 24.03\% \\
A16	& 0.6671 & 0.8428 & 0.8961 & 34.33\% &  6.32\% \\
A17	& 1.0000 & 1.0000 & 1.0000 &  0.00\% &  0.00\% \\
A18	& 0.8465 &      / & 0.8540 &  0.89\% &       / \\
A19	& 0.6906 & 0.9561 & 0.9583 & 38.76\% &  0.23\% \\
A20	& 0.8020 & 0.8708 & 0.8803 &  9.76\% &  1.09\% \\ \midrule
Average & 0.7645 & 0.8527 & 0.8877 &  4.10\% & 16.12\% \\ \bottomrule
\end{tabular}}
\end{table}

According to Table \ref{tab:rq3}, LLMPrior achieves an improvement of 4.1\% on average over the DirectLLMPrior (denoted as L - D in Table \ref{tab:rq3}) and 16.12\% on average over the SimpleLLMPrior (denoted as L - S in Table \ref{tab:rq3}). Overall LLMPrior is able to further enhance the prioritization effect comparing to the DirectLLMPrior and SimpleLLMPrior.

We perform Wilcoxon signed-rank tests to assess the significance of LLMPrior's performance enhancement over SimpleLLMPrior and DirectLLMPrior. The test yields p-values of $1.89\times 10^{-3}$ and $5.21\times 10^{-3}$ for the comparisons with SimpleLLMPrior and DirectLLMPrior, respectively. Given that both p-values are below the widely accepted significance threshold of 0.05, we can conclude that the improvements offered by LLMPrior are statistically significant. This underscores the efficacy of the cluster-based prioritization strategy and prompt engineering techniques implemented in LLMPrior.

In Figure \ref{fig:rq3}, we display the APFD value distribution according to the result of each repetition. LLMPrior shows better stability at multiple repetitions, and compared to the alternative approaches, the data distribution is more centralized, and the performance is less affected by the randomness of the LLM-generated content.

The significant improvement by comparing SimpleLLMPrior with LLMPrior and DirectLLMPrior, especially on apps with more reports, demonstrates the importance of prompt engineering techniques when interacting with LLMs. Moreover, by comparing LLMPrior with DirectLLMPrior, we summerize how the cluster-based prioritization strategy contributes to the better effectiveness of LLMPrior.

Firstly, LLM's behavior in the prioritizing reports is of certain unreasonable randomness. For example, we observe that in some of the prioritization results LLM ranks two reports of the same bug type at the top of the sequence. According to LLM's explanation, LLM is influenced by the common understanding of the word ``prioritize'', and tends to rank bugs that it considers to be of higher severity at the front of the sequence even if repeated. This causes the prioritization results to exhibit a decrease in effectiveness. The cluster-based prioritization strategy circumvents this randomness by using algorithm instead of LLM in prioritizing test reports after report clustering.

Additionally, when analyzing the reports, DirectLLMPrior simply categorizes all reports into several parallel categories. Due to the uncertainty of the content generated by the LLM, its classification criteria are not completely consistent each time. When its categorization results are inaccurate or the categorization granularity is not properly selected, the prioritization effect will be significantly reduced. This will lead to large fluctuations in the effect, which is shown in the box-and-line diagram as longer boxes, larger range between maximum and minimum values, and more outliers. LLMPrior better handles the problem of categorization granularity by guiding the LLM to cluster the reports with a multi-layered tree structure, which makes the prioritization results more reliable, robust, and stable.

Last but not least, LLMPrior surpasses other approaches by effectively processing numerous reports simultaneously. Unlike DirectLLMPrior and SimpleLLMPrior, which struggle to prioritize reports when dealing with apps that have over 100 reports, LLMPrior demonstrates superior capability. This distinction highlights LLMPrior's enhanced stability, reliability, and performance, marking a significant improvement over the two alternative approaches.

\subsection{Answer to RQ4}

We also assess the token efficiency of LLMPrior by calculating the token usage during the repeated crowdsourced test report prioritization process.

As detailed in Table \ref{tab:rq4}, the average TPR of LLMPrior is around 93, with a standard deviation of 41.13. Our analysis reveals that LLMPrior on average uses 42.7 more tokens per report than SimpleLLMPrior. However, considering LLMPrior far surpasses SimpleLLMPrior in effectiveness, this extra cost is acceptable. Compared to DirectLLMPrior, LLMPrior is more token-efficient. On average, LLMPrior saves 53.3 tokens on each report, with a range of 20 to 101 tokens across different apps. Furthermore, LLMPrior has a lower standard deviation (41.13 v.s. 55.25) on TPR, indicating that its costs are more stable. This efficiency stems from LLMPrior's shorter prompts and responses, focusing solely on report clustering, thereby reducing the token usage required for analysis and prioritization.

Similarly, we conduct the Wilcoxon signed-rank test to evaluate the token efficiency improvement achieved by LLMPrior over DirectLLMPrior. The calculated p-value is $3.05\times 10^{-5}$, far less than the preset threshold 0.05, which means the token saving is statistical significant. Furthermore, the calculated effect size value (Cohen's d) is 1.09, which exceeds 0.8 and suggests that the token savings are practically significant as well. Consequently, LLMPrior emerges as a more token efficient approach compared to DirectLLMPrior. It has the potential to deal with larger crowdsourced test report dataset.

However, it is important to note that while LLMPrior offers certain advantages in token efficiency, its large-scale application will inevitably require significant computing resources due its LLM-based feature. Consequently, developers must consider the available computing resources and their expectations for the efficiency of reviewing test reports, making a compromise based on the actual circumstances.

\begin{table}[!h]
\caption{LLMPrior Report Prioritization Efficiency and Comparison with LLM Baselines}
\label{tab:rq4}
\centering
\scalebox{0.8}{
\renewcommand{\arraystretch}{1.1}
\begin{tabular}{c|rr|r|rr}
\toprule
\multirow{2}{*}{App} & \multicolumn{3}{c|}{TPR} & \multicolumn{2}{c}{Difference} \\ \cline{2-6}
&
\multicolumn{1}{c}{Simple} &
\multicolumn{1}{c|}{Direct} &
\multicolumn{1}{c|}{LLMPrior} &
\multicolumn{1}{c}{S - L} &
\multicolumn{1}{c}{D - L} \\ \midrule
A3	    & 46.21 &  118.43 &  69.95 &  -23.74 &  48.48 \\
A4	    & 45.89 &  165.01 & 107.56 &  -61.67 &  57.45 \\
A5	    & 43.32 &  148.20 &  94.16 &  -50.84 &  54.04 \\
A6	    & 50.65 &  135.52 &  72.21 &  -21.56 &  63.31 \\
A8	    & 38.47 &   80.70 &  52.79 &  -14.32 &  27.91 \\
A9	    & 52.83 &   86.04 &  66.00 &  -13.17 &  20.04 \\
A10	    & 44.58 &  134.44 &  71.21 &  -26.63 &  63.23 \\
A11	    & 63.60 &  238.67 & 182.27 & -118.67 &  56.40 \\
A12	    & 43.50 &  211.60 & 110.60 &  -67.10 & 101.00 \\
A13	    & 49.62 &  153.94 &  87.94 &  -38.32 &  66.00 \\
A15	    & 49.72 &   78.40 &  58.70 &   -8.98 &  19.70 \\
A16	    & 49.44 &  103.30 &  64.81 &  -15.37 &  38.49 \\
A17	    & 59.75 &  253.86 & 188.17 & -128.42 &  65.69 \\
A19	    & 76.85 &  182.72 & 100.31 &  -23.46 &  82.41 \\
A20	    & 47.94 &  111.55 &  75.74 &  -27.80 &  35.81 \\ \midrule
AVG     & 50.82 &  146.83 &  93.49 &  -42.67 &  53.34 \\ \midrule
S.D.    &  9.58 &   55.25 &  41.13 &   37.20 &  22.51 \\ \bottomrule
\end{tabular}}
\end{table}

\section{Discussion}

\textbf{Trade-offs in practical use.} LLMPrior, which utilizes the capabilities of LLMs, introduces some trade-offs in practical use when compared to traditional NLP models. First, it generally has a longer runtime. Our estimates suggest that LLMPrior takes approximately 0.2 seconds per report during prioritization, which is nearly twice as long as traditional NLP models that do not use GPU acceleration. Second, LLMPrior incurs additional costs. Based on OpenAI’s API pricing and the TPR data we've collected, the estimated cost per report during prioritization is around \$0.0003. Overall, while LLMPrior does introduce extra time and financial costs, these are within an acceptable range. Importantly, as demonstrated in Section 4.2, LLMPrior significantly improves prioritization quality compared to traditional NLP models, offering a unique advantage.

\textbf{The complementary role of fault detection}. The evaluation of crowdsourced test report prioritization approaches has primarily focused on statistical assessments of prioritization results, such as calculating and analyzing APFD metrics. However, for practical engineering applications, incorporating fault detection could provide a more robust evaluation of crowdsourced test report prioritization. In this process, prioritized test reports are submitted to reviewers who verify the reported issues and communicate them to the developers. By analyzing this end-to-end process, we can gain deeper insights into the real-world effectiveness and value of the prioritization approaches. Future research could involve case studies on fault detection workflows as a means to achieve a more comprehensive evaluation of crowdsourced test report prioritization approaches.

\section{Threats to Validity}

The enrollment of crowdworkers in taking the testing tasks is uncontrolled, leading to potential discrepancies in report quality. Some reports with low quality may include vague linguistic descriptions or ambiguous screenshots, which could mislead the LLM. When the LLM analyzes these reports, it might incorrectly categorize and prioritize them due to the hallucination phenomenon of LLM. If these reports are deemed to reveal unique types of bugs, they have a probability of receiving a high priority. Despite this, it is important to note that such low-quality reports minimally impact the overall prioritization process and result. Moreover, the developers will manually review these reports afterwards. The presence of a small number of poorly written reports will only slightly increase the inspection workload, which is acceptable to some extent.

Our dataset is exclusively composed of Chinese-language content, which poses a potential threat. However, large language models, renowned for exceptional generalizability, demonstrate remarkable proficiency in understanding, analyzing, and translating content in various languages. Leveraging this capability, we anticipate that the LLM will effectively prioritize crowdsourced test reports, regardless of their linguistic composition.

In our study, we gathered 1,417 reports across 20 apps. However, the distribution of these reports is uneven across the apps. Certain apps have fewer than 10 reports each, presenting a scenario where the need for automated techniques to aid developers is not typical. To mitigate risks, our focus is primarily on the outcomes of experiments with a relative large number of reports.

We use only one LLM as a representative. We typically use OpenAI's GPT-4-Turbo, which is a closed-sourced model so far, as a representative of LLM in our study. Other closed-sourced LLMs like Claude or open-sourced LLMs like LLaMA and ChatGLM are not involved. Additionally, as the model we use is close-sourced, we do not have the ability to further evaluate how fine-tuning techniques will have influence on the test report prioritization task. We predict that the effect will be negatively impacted because GPT-4-Turbo is the most advanced LLM we have ever known. However, the difference lies in the inherent comprehension and reasoning abilities of different LLMs and is not relevant to the ideas and design of our approach.

\section{Related Work}

\subsection{Crowdsourced Test Report Processing}

The process of crowdsourced test reports aims at better manual review efficiency for app developers. There is a diverse range of approaches that researches employ in their studies on this subject.

Clustering and classification are central themes in some research. Wang et al. develop LOAF \cite{wang_local-based_2016} to classify true fault from test reports using textual features. They also build a cluster-based classifier \cite{wang_towards_2016} that applies a k-means algorithm for effective fault discrimination and use a deep learning model \cite{wang_domain_2017} for cross-domain classification by discovering intermediate representation across domains. Jiang et al. \cite{jiang_fuzzy_2018} introduce TERFUR for clustering redundant and multi-bug reports, easing the burden on developers. Hao et al. \cite{hao_ctras_2019} create CTRAS, which leverages duplicates to flesh out bug descriptions by combining text and screenshots. Similarly, Cai et al. \cite{cai_reports_2021} utilize  fusing features from text, bug type, and screenshots to cluster crowdsourced test reports. Chen et al. \cite{chen_effective_2021} propose RCSE, a model encoding test report description information at sentence level to cluster similar reports. Liu et al. \cite{liu_clustering_2022} measures screenshot similarity with Spatial Pyramid Matching and text distance with NLP. Li et al. \cite{li_classifying_2022} extract features from texts and images, and then conduct an empirical study on six widely-used classification algorithms with their feature fusion approach. Li et al. \cite{li_identifying_2022} propose FER to counter low-quality and imbalanced distributions of bug reports relative to their severity. Du et al. \cite{du_semcluster_2022} present a tool named SemCluster, which makes the most of the semantic connection between textual descriptions and screenshots by constructing semantic binding rules and performing semi-supervised test report clustering.

Research has also been directed at detecting duplicate reports, an approach predating crowdsourced testing. Runeson et al. \cite{runeson_detection_2007} first propose an approach using NLP techniques to assess report similarity. Sun et al. \cite{sun_discriminative_2010} \cite{sun_towards_2011} calculate the textual similarity of bug reports with a discriminative model and information retrieval functions. Alipour et al. \cite{alipour_contextual_2013} take domain-specific context into account to improve information-retrieval methods for bug duplication. While Lazar et al. \cite{lazar_improving_2014} improve the accuracy of duplicate bug report detection with new textual features. Ebrahimi et al. \cite{ebrahimi_hmm-based_2019} employ Hidden Markov Models and stack traces to detects duplicate reports. SETU \cite{wang_images_2019} proposed by Wang et al. combines screenshots with texts for feature extraction. Then report similarities are derived from the features with a weighed summary. Kim and Yang \cite{kim_predicting_2022} predict redundancy in bug report using the BERT algorithm and topic-based duplicate/non-duplicate feature extraction. Nguyen and Nguyen \cite{nguyen_incremental_2022} utilize Relational Topic Model (RTM) to formulate the probabilistic structures of technical aspects in a collection of bug reports and the duplication indicators among them. Wu et al. \cite{wu_intelligent_2023} propose CTEDB, a new method for detecting duplicate bug reports based on technical term extraction,

The approaches above choose to select only a subset of the test reports as representative, while another point of view is that all the reports contain valuable information despite the presence of redundancy. Developers still have to inspect most reports to carry on the subsequent bug-fixing work. This leads to the propose of crowdsourced test report prioritization, which offers an optimized sequence of all the test reports to app developers.

DivRisk \cite{feng_test_2015}, the first proposed test report prioritization technique, utilizes two strategies to identify diverse reports that are likely to reveal faults. Yang et al. \cite{yang_crowdsourced_2022} propose DivClass which combines NLP with diversity and classification strategies for text-based report prioritization. Feng et al. \cite{feng_multi-objective_2016} employ the Spatial Pyramid Matching technique to analyze screenshots in the reports and apply NLP techniques to process textual descriptions. Tong and Zhang \cite{tong_crowdsourced_2021} introduce bug severity to the considering factors of prioritization. They employ hashing in their prioritization algorithm, after feature extraction from textual descriptions and screenshots. DeepPrior \cite{yu_prioritize_2021} extracts features from app screenshots and textual descriptions separately and aggregates them before the prioritization.

The aforementioned studies all engage in extracting text features from crowdsourced test reports for optimization purposes. Despite the varied techniques employed across these approaches, the processing of texts remains a critical element. This observation motivates us to devise an approach that offers a more profound comprehension of textual descriptions in the crowdsourced test reports.

\subsection{Large Language Model for Software Engineering}

Recent advancements in artificial intelligence have led to significant breakthroughs in large language models (LLMs), with rapid evolution and increasing applicability in various fields, notably in software engineering research.

In the realm of software testing practices, several studies exemplify the innovative applications of LLMs. From the nuanced task of generating semantic input text for GUI testing, as explored by Liu et al. \cite{liu_fill_2023}, to the more complex automation of test generation from bug reports by Kang et al. \cite{kang_large_2023}, LLMs have shown remarkable versatility. Their capability extends to enhancing GUI test exploration through human-like interactions, a concept brought to life by Liu et al. \cite{liu_chatting_2023}, and to serving as automated testing assistants, a blueprint of which is discussed by Feldt et al. \cite{feldt_towards_2023}. Additinally, Yu et al. \cite{yu_llm_2023} explore the potential of automated test script generation and migration with LLMs.

The impact of LLMs is equally profound at the code level. MacNeil et al. \cite{macneil_generating_2022} focus on LLMs' potential to support learning by explaining numerous aspects of a given code snippet. Moving beyond traditional methodologies, Siddiq et al. \cite{siddiq_zero-shot_2023} show how LLMs can predict code complexity and conduct an empirical evaluation. The study by Mashhadi et al. \cite{mashhadi_method-level_2023} on using a pre-trained CodeBERT to predict bug severity exemplifies their analytical ability. Moreover, the evaluation of LLMs in detecting software vulnerabilities by Purba et al. \cite{purba_software_2023} and the LLM-based program repair technique proposed by Weng and Andrzejak \cite{weng_automatic_2023}, highlight the models' potential in enhancing software reliability and security.

Beyond these specific applications, LLMs also find their place in broader domains. Petrović et al.\cite{petrovic_chatgpt_2023} explore how ChatGPT can be used within IoT systems, while Ding et al. \cite{ding_hpc-gpt_2023} fine-tune a LLM for better performance in high-performance computing (HPC) domain tasks. Koreeda et al. \cite{koreeda_larch_2023} develop LARCH, combining representative code identification techniques with LLM to generate coherent and factually correct code repository documentation.

All the above studies effectively integrate LLMs into software engineering practices. They utilize the remarkable proficiency of LLMs in comprehending, analyzing, and generating both natural language and code. This integration aids in resolving issues and refining methodologies within software engineering. This inspires us to apply LLM's capabilities in prioritizing crowdsourced test reports. We aim to leverage LLMs for in-depth semantic understanding of text descriptions in test reports and enhance crowdsourced test report prioritization with more effective and reliable results.

\section{Conclusion}

In this paper, we propose LLMPrior, the first crowdsourced test report prioritization approach that utilizes a large language model. Instead of directly guiding LLM to prioritize the test reports, we design a cluster-based prioritization strategy, which can guarantee the stable generation of complete prioritization results. Following certain prompt engineering techniques, we guide the LLM to cluster the reports by the bug types revealed in their textual descriptions and organize the clustering result in a multi-way tree structure. We then convert the LLM output into a hierarchical cluster tree. Finally, we implement a recurrent selection algorithm to generate prioritized test report sequence based on the tree. We conduct an empirical experiment and the results show that LLMPrior outperforms the state-of-the-art approach DeepPrior by 12.77\% on average and is superior in efficiency and effectiveness with the help of prompt engineering techniques and the cluster-based prioritization strategy.

\section*{Acknowledgment}

We would like to thank the anonymous reviewers for their insightful comments. This work is supported partially by the National Natural Science Foundation of China (62372228) and the Fundamental Research Funds for the Central Universities (14380029).

\bibliographystyle{elsarticle-num} 
\bibliography{main}

\begin{thebibliography}{10}
\expandafter\ifx\csname url\endcsname\relax
  \def\url#1{\texttt{#1}}\fi
\expandafter\ifx\csname urlprefix\endcsname\relax\def\urlprefix{URL }\fi
\expandafter\ifx\csname href\endcsname\relax
  \def\href#1#2{#2} \def\path#1{#1}\fi

\bibitem{wei_taming_2016}
L.~Wei, Y.~Liu, S.-C. Cheung, Taming android fragmentation: characterizing and detecting compatibility issues for android apps, in: Proceedings of the 31st {IEEE}/{ACM} International Conference on Automated Software Engineering, {ASE} '16, pp. 226--237.
\newblock \href {https://doi.org/10.1145/2970276.2970312} {\path{doi:10.1145/2970276.2970312}}.

\bibitem{zhang_crowdsourced_2017}
T.~Zhang, J.~Gao, J.~Cheng, Crowdsourced testing services for mobile apps, in: 2017 {IEEE} Symposium on Service-Oriented System Engineering ({SOSE}), 2017, pp. 75--80.
\newblock \href {https://doi.org/10.1109/SOSE.2017.28} {\path{doi:10.1109/SOSE.2017.28}}.

\bibitem{yu_lirat_2019}
S.~Yu, C.~Fang, Y.~Feng, W.~Zhao, Z.~Chen, {LIRAT}: {Layout} and {Image} {Recognition} {Driving} {Automated} {Mobile} {Testing} of {Cross}-{Platform}, in: 2019 34th {IEEE}/{ACM} {International} {Conference} on {Automated} {Software} {Engineering} ({ASE}), 2019, pp. 1066--1069.
\newblock \href {https://doi.org/10.1109/ASE.2019.00103} {\path{doi:10.1109/ASE.2019.00103}}.

\bibitem{gao_successes_2019}
R.~Gao, Y.~Wang, Y.~Feng, Z.~Chen, W.~Eric~Wong, Successes, challenges, and rethinking – an industrial investigation on crowdsourced mobile application testing 24~(2) (2019) 537--561.
\newblock \href {https://doi.org/10.1007/s10664-018-9618-5} {\path{doi:10.1007/s10664-018-9618-5}}.

\bibitem{wang_intelligent_2022}
Q.~Wang, Z.~Chen, J.~Wang, Y.~Feng, Intelligent Crowdsourced Testing, Springer Nature Singapore, 2022.
\newblock \href {https://doi.org/10.1007/978-981-16-9643-5} {\path{doi:10.1007/978-981-16-9643-5}}.

\bibitem{yu_prioritize_2021}
S.~Yu, C.~Fang, Z.~Cao, X.~Wang, T.~Li, Z.~Chen, Prioritize crowdsourced test reports via deep screenshot understanding, in: 2021 {IEEE}/{ACM} 43rd {International} {Conference} on {Software} {Engineering} ({ICSE}), 2021, pp. 946--956.
\newblock \href {https://doi.org/10.1109/ICSE43902.2021.00090} {\path{doi:10.1109/ICSE43902.2021.00090}}.

\bibitem{wang_images_2019}
J.~Wang, M.~Li, S.~Wang, T.~Menzies, Q.~Wang, Images don’t lie: {Duplicate} crowdtesting reports detection with screenshot information, Information and Software Technology 110 (2019) 139--155.
\newblock \href {https://doi.org/10.1016/j.infsof.2019.03.003} {\path{doi:10.1016/j.infsof.2019.03.003}}.

\bibitem{feng_test_2015}
Y.~Feng, Z.~Chen, J.~A. Jones, C.~Fang, B.~Xu, Test report prioritization to assist crowdsourced testing, in: Proceedings of the 2015 10th {Joint} {Meeting} on {Foundations} of {Software} {Engineering}, {ESEC}/{FSE} 2015, 2015, pp. 225--236.
\newblock \href {https://doi.org/10.1145/2786805.2786862} {\path{doi:10.1145/2786805.2786862}}.

\bibitem{feng_multi-objective_2016}
Y.~Feng, J.~A. Jones, Z.~Chen, C.~Fang, Multi-objective test report prioritization using image understanding, in: 2016 31st {IEEE}/{ACM} {International} {Conference} on {Automated} {Software} {Engineering} ({ASE}), 2016, pp. 202--213.

\bibitem{liu_clustering_2022}
D.~Liu, Y.~Feng, X.~Zhang, J.~A. Jones, Z.~Chen, Clustering crowdsourced test reports of mobile applications using image understanding, IEEE Transactions on Software Engineering 48~(4) (2022) 1290--1308.
\newblock \href {https://doi.org/10.1109/TSE.2020.3017514} {\path{doi:10.1109/TSE.2020.3017514}}.

\bibitem{li_classifying_2022}
Y.~Li, Y.~Feng, R.~Hao, D.~Liu, C.~Fang, Z.~Chen, B.~Xu, Classifying crowdsourced mobile test reports with image features: An empirical study, J. Syst. Softw. 184~(C) (2022).
\newblock \href {https://doi.org/10.1016/j.jss.2021.111121} {\path{doi:10.1016/j.jss.2021.111121}}.

\bibitem{li_identifying_2022}
H.~Li, X.~Qi, M.~Li, Y.~Qu, X.~Ge, Identifying high-impact bug reports with imbalance distribution by instance fuzzy entropy, International Journal of Software Engineering and Knowledge Engineering 32~(09) (2022) 1389--1417.
\newblock \href {https://doi.org/10.1142/S021819402250053X} {\path{doi:10.1142/S021819402250053X}}.

\bibitem{du_semcluster_2022}
M.~Du, S.~Yu, C.~Fang, T.~Li, H.~Zhang, Z.~Chen, {SemCluster}: a semi-supervised clustering tool for crowdsourced test reports with deep image understanding, in: Proceedings of the 30th {ACM} {Joint} {European} {Software} {Engineering} {Conference} and {Symposium} on the {Foundations} of {Software} {Engineering}, {ESEC}/{FSE} 2022, 2022, pp. 1756--1759.
\newblock \href {https://doi.org/10.1145/3540250.3558933} {\path{doi:10.1145/3540250.3558933}}.

\bibitem{kim_predicting_2022}
T.~Kim, G.~Yang, Predicting {Duplicate} in {Bug} {Report} {Using} {Topic}-{Based} {Duplicate} {Learning} {With} {Fine} {Tuning}-{Based} {BERT} {Algorithm}, IEEE Access 10 (2022) 129666--129675.
\newblock \href {https://doi.org/10.1109/ACCESS.2022.3226238} {\path{doi:10.1109/ACCESS.2022.3226238}}.

\bibitem{nguyen_incremental_2022}
A.~T. Nguyen, T.~N. Nguyen, Incremental {Relational} {Topic} {Model} for {Duplicate} {Bug} {Report} {Detection}, in: 2022 29th {Asia}-{Pacific} {Software} {Engineering} {Conference} ({APSEC}), 2022, pp. 99--108.
\newblock \href {https://doi.org/10.1109/APSEC57359.2022.00022} {\path{doi:10.1109/APSEC57359.2022.00022}}.

\bibitem{wu_intelligent_2023}
X.~Wu, W.~Shan, W.~Zheng, Z.~Chen, T.~Ren, X.~Sun, An {Intelligent} {Duplicate} {Bug} {Report} {Detection} {Method} {Based} on {Technical} {Term} {Extraction}, in: 2023 {IEEE}/{ACM} {International} {Conference} on {Automation} of {Software} {Test} ({AST}), 2023, pp. 1--12.
\newblock \href {https://doi.org/10.1109/AST58925.2023.00005} {\path{doi:10.1109/AST58925.2023.00005}}.

\bibitem{tong_crowdsourced_2021}
Y.~Tong, X.~Zhang, Crowdsourced test report prioritization considering bug severity, Information and Software Technology 139 (2021) 106668.
\newblock \href {https://doi.org/10.1016/j.infsof.2021.106668} {\path{doi:10.1016/j.infsof.2021.106668}}.

\bibitem{NIPS2013_9aa42b31}
T.~Mikolov, I.~Sutskever, K.~Chen, G.~S. Corrado, J.~Dean, Distributed representations of words and phrases and their compositionality, in: Advances in Neural Information Processing Systems, Vol.~26, 2013.

\bibitem{shen_comparison_2021}
Y.~Shen, J.~Liu, Comparison of text sentiment analysis based on bert and word2vec, in: 2021 {IEEE} 3rd International Conference on Frontiers Technology of Information and Computer ({ICFTIC}), pp. 144--147.
\newblock \href {https://doi.org/10.1109/ICFTIC54370.2021.9647258} {\path{doi:10.1109/ICFTIC54370.2021.9647258}}.

\bibitem{huang_survey_2020}
S.~Huang, H.~Chen, Z.~Hui, Y.~Liu, A survey of the use of test report in crowdsourced testing, in: 2020 {IEEE} 20th International Conference on Software Quality, Reliability and Security ({QRS}), pp. 430--441.
\newblock \href {https://doi.org/10.1109/QRS51102.2020.00062} {\path{doi:10.1109/QRS51102.2020.00062}}.

\bibitem{tamkin_understanding_2021}
A.~Tamkin, M.~Brundage, J.~Clark, D.~Ganguli, Understanding the capabilities, limitations, and societal impact of large language models.
\newblock \href {https://doi.org/10.48550/arXiv.2102.02503} {\path{doi:10.48550/arXiv.2102.02503}}.

\bibitem{wei_emergent_2022}
J.~Wei, Y.~Tay, R.~Bommasani, C.~Raffel, B.~Zoph, S.~Borgeaud, D.~Yogatama, M.~Bosma, D.~Zhou, D.~Metzler, E.~H. Chi, T.~Hashimoto, O.~Vinyals, P.~Liang, J.~Dean, W.~Fedus, Emergent abilities of large language models (2022).
\newblock \href {https://doi.org/10.48550/arXiv.2206.07682} {\path{doi:10.48550/arXiv.2206.07682}}.

\bibitem{chang_survey_2023}
Y.~Chang, X.~Wang, J.~Wang, Y.~Wu, L.~Yang, K.~Zhu, H.~Chen, X.~Yi, C.~Wang, Y.~Wang, W.~Ye, Y.~Zhang, Y.~Chang, P.~S. Yu, Q.~Yang, X.~Xie, A survey on evaluation of large language models, ACM Trans. Intell. Syst. Technol. 15~(3) (2024).
\newblock \href {https://doi.org/10.1145/3641289} {\path{doi:10.1145/3641289}}.

\bibitem{hou_large_2023}
X.~Hou, Y.~Zhao, Y.~Liu, Z.~Yang, K.~Wang, L.~Li, X.~Luo, D.~Lo, J.~Grundy, H.~Wang, Large language models for software engineering: A systematic literature review.
\newblock \href {https://doi.org/10.48550/arXiv.2308.10620} {\path{doi:10.48550/arXiv.2308.10620}}.

\bibitem{zheng_towards_2023}
Z.~Zheng, K.~Ning, J.~Chen, Y.~Wang, W.~Chen, L.~Guo, W.~Wang, Towards an understanding of large language models in software engineering tasks.
\newblock \href {https://doi.org/10.48550/arXiv.2308.11396} {\path{doi:10.48550/arXiv.2308.11396}}.

\bibitem{belzner_large_2024}
L.~Belzner, T.~Gabor, M.~Wirsing, Large language model assisted software engineering: Prospects, challenges, and a case study, in: Bridging the Gap Between {AI} and Reality, Lecture Notes in Computer Science, pp. 355--374.
\newblock \href {https://doi.org/10.1007/978-3-031-46002-9\_23} {\path{doi:10.1007/978-3-031-46002-9\_23}}.

\bibitem{wei_chain--thought_2023}
J.~Wei, X.~Wang, D.~Schuurmans, M.~Bosma, B.~Ichter, F.~Xia, E.~H. Chi, Q.~V. Le, D.~Zhou, Chain-of-thought prompting elicits reasoning in large language models, in: Proceedings of the 36th International Conference on Neural Information Processing Systems, NIPS '22, 2024.

\bibitem{kojima_large_2022}
T.~Kojima, S.~S. Gu, M.~Reid, Y.~Matsuo, Y.~Iwasawa, Large language models are zero-shot reasoners, in: Proceedings of the 36th International Conference on Neural Information Processing Systems, NIPS '22, 2024.

\bibitem{zhou_large_2023}
Y.~Zhou, A.~I. Muresanu, Z.~Han, K.~Paster, S.~Pitis, H.~Chan, J.~Ba, Large {Language} {Models} {Are} {Human}-{Level} {Prompt} {Engineers} (2023).
\newblock \href {https://doi.org/10.48550/arXiv.2211.01910} {\path{doi:10.48550/arXiv.2211.01910}}.

\bibitem{mao_survey_2017}
K.~Mao, L.~Capra, M.~Harman, Y.~Jia, A survey of the use of crowdsourcing in software engineering 126  57--84.
\newblock \href {https://doi.org/10.1016/j.jss.2016.09.015} {\path{doi:10.1016/j.jss.2016.09.015}}.

\bibitem{yu_mobile_2023}
S.~Yu, C.~Fang, Q.~Zhang, Z.~Cao, Y.~Yun, Z.~Cao, K.~Mei, Z.~Chen, Mobile app crowdsourced test report consistency detection via deep image-and-text fusion understanding  1--20\href {https://doi.org/10.1109/TSE.2023.3285787} {\path{doi:10.1109/TSE.2023.3285787}}.

\bibitem{mao_developer_2015}
K.~Mao, Y.~Yang, Q.~Wang, Y.~Jia, M.~Harman, Developer recommendation for crowdsourced software development tasks, in: 2015 {IEEE} Symposium on Service-Oriented System Engineering, pp. 347--356.
\newblock \href {https://doi.org/10.1109/SOSE.2015.46} {\path{doi:10.1109/SOSE.2015.46}}.

\bibitem{wang_isense_2019}
J.~Wang, Y.~Yang, R.~Krishna, T.~Menzies, Q.~Wang, {iSENSE}: Completion-aware crowdtesting management, in: 2019 {IEEE}/{ACM} 41st International Conference on Software Engineering ({ICSE}), pp. 912--923.
\newblock \href {https://doi.org/10.1109/ICSE.2019.00097} {\path{doi:10.1109/ICSE.2019.00097}}.

\bibitem{wang_isense20_2020}
J.~Wang, Y.~Yang, T.~Menzies, Q.~Wang, {iSENSE}2.0: Improving completion-aware crowdtesting management with duplicate tagger and sanity checker 29~(4)  24:1--24:27.
\newblock \href {https://doi.org/10.1145/3394602} {\path{doi:10.1145/3394602}}.

\bibitem{wang_context-aware_2022}
J.~Wang, Y.~Yang, S.~Wang, C.~Chen, D.~Wang, Q.~Wang, Context-aware personalized crowdtesting task recommendation 48~(8)  3131--3144.
\newblock \href {https://doi.org/10.1109/TSE.2021.3081171} {\path{doi:10.1109/TSE.2021.3081171}}.

\bibitem{Wang2022}
Q.~Wang, Z.~Chen, J.~Wang, Y.~Feng, Crowdsourced Testing Task Management, Springer Nature Singapore, 2022, pp. 91--122.
\newblock \href {https://doi.org/10.1007/978-981-16-9643-5\_7} {\path{doi:10.1007/978-981-16-9643-5\_7}}.

\bibitem{yang_crowdsourced_2022}
Y.~Yang, X.~Chen, Crowdsourced {Test} {Report} {Prioritization} {Based} on {Text} {Classification}, IEEE Access 10 (2022) 92692--92705.
\newblock \href {https://doi.org/10.1109/ACCESS.2021.3128726} {\path{doi:10.1109/ACCESS.2021.3128726}}.

\bibitem{brown_language_2020}
T.~B. Brown, B.~Mann, N.~Ryder, M.~Subbiah, J.~Kaplan, P.~Dhariwal, A.~Neelakantan, P.~Shyam, G.~Sastry, A.~Askell, S.~Agarwal, A.~Herbert-Voss, G.~Krueger, T.~Henighan, R.~Child, A.~Ramesh, D.~M. Ziegler, J.~Wu, C.~Winter, C.~Hesse, M.~Chen, E.~Sigler, M.~Litwin, S.~Gray, B.~Chess, J.~Clark, C.~Berner, S.~McCandlish, A.~Radford, I.~Sutskever, D.~Amodei, Language {Models} are {Few}-{Shot} {Learners} (2020).
\newblock \href {https://doi.org/10.48550/arXiv.2005.14165} {\path{doi:10.48550/arXiv.2005.14165}}.

\bibitem{rothermel_prioritizing_2001}
G.~Rothermel, R.~Untch, C.~Chu, M.~Harrold, Prioritizing test cases for regression testing, IEEE Transactions on Software Engineering 27~(10) (2001) 929--948.
\newblock \href {https://doi.org/10.1109/32.962562} {\path{doi:10.1109/32.962562}}.

\bibitem{wang_local-based_2016}
J.~Wang, S.~Wang, Q.~Cui, Q.~Wang, Local-based active classification of test report to assist crowdsourced testing, in: 2016 31st IEEE/ACM International Conference on Automated Software Engineering (ASE), 2016, pp. 190--201.
\newblock \href {https://doi.org/10.1145/2970276.2970300} {\path{doi:10.1145/2970276.2970300}}.

\bibitem{wang_towards_2016}
J.~Wang, Q.~Cui, Q.~Wang, S.~Wang, Towards {Effectively} {Test} {Report} {Classification} to {Assist} {Crowdsourced} {Testing}, in: Proceedings of the 10th {ACM}/{IEEE} {International} {Symposium} on {Empirical} {Software} {Engineering} and {Measurement}, Ciudad Real Spain, 2016, pp. 1--10.
\newblock \href {https://doi.org/10.1145/2961111.2962584} {\path{doi:10.1145/2961111.2962584}}.

\bibitem{wang_domain_2017}
J.~Wang, Q.~Cui, S.~Wang, Q.~Wang, Domain adaptation for test report classification in crowdsourced testing, in: 2017 {IEEE}/{ACM} 39th International Conference on Software Engineering: Software Engineering in Practice Track ({ICSE}-{SEIP}), 2017, pp. 83--92.
\newblock \href {https://doi.org/10.1109/ICSE-SEIP.2017.8} {\path{doi:10.1109/ICSE-SEIP.2017.8}}.

\bibitem{jiang_fuzzy_2018}
H.~Jiang, X.~Chen, T.~He, Z.~Chen, X.~Li, Fuzzy {Clustering} of {Crowdsourced} {Test} {Reports} for {Apps}, ACM Transactions on Internet Technology 18~(2) (2018) 18:1--18:28.
\newblock \href {https://doi.org/10.1145/3106164} {\path{doi:10.1145/3106164}}.

\bibitem{hao_ctras_2019}
R.~Hao, Y.~Feng, J.~A. Jones, Y.~Li, Z.~Chen, {CTRAS}: {Crowdsourced} {Test} {Report} {Aggregation} and {Summarization}, in: 2019 {IEEE}/{ACM} 41st {International} {Conference} on {Software} {Engineering} ({ICSE}), 2019, pp. 900--911.
\newblock \href {https://doi.org/10.1109/ICSE.2019.00096} {\path{doi:10.1109/ICSE.2019.00096}}.

\bibitem{cai_reports_2021}
L.~Cai, N.~Wang, M.~Chen, J.~Wang, J.~Wang, J.~Gong, Reports aggregation of crowdsourcing test based on feature fusion, in: 2021 IEEE 21st International Conference on Software Quality, Reliability and Security Companion (QRS-C), 2021, pp. 51--59.
\newblock \href {https://doi.org/10.1109/QRS-C55045.2021.00018} {\path{doi:10.1109/QRS-C55045.2021.00018}}.

\bibitem{chen_effective_2021}
H.~Chen, S.~Huang, Y.~Liu, R.~Luo, Y.~Xie, An effective crowdsourced test report clustering model based on sentence embedding, in: 2021 IEEE 21st International Conference on Software Quality, Reliability and Security (QRS), 2021, pp. 888--899.
\newblock \href {https://doi.org/10.1109/QRS54544.2021.00098} {\path{doi:10.1109/QRS54544.2021.00098}}.

\bibitem{runeson_detection_2007}
P.~Runeson, M.~Alexandersson, O.~Nyholm, Detection of duplicate defect reports using natural language processing, in: 29th International Conference on Software Engineering ({ICSE}'07), 2007, pp. 499--510.
\newblock \href {https://doi.org/10.1109/ICSE.2007.32} {\path{doi:10.1109/ICSE.2007.32}}.

\bibitem{sun_discriminative_2010}
C.~Sun, D.~Lo, X.~Wang, J.~Jiang, S.-C. Khoo, A discriminative model approach for accurate duplicate bug report retrieval, in: Proceedings of the 32nd {ACM}/{IEEE} International Conference on Software Engineering - Volume 1, {ICSE} '10, 2010, pp. 45--54.
\newblock \href {https://doi.org/10.1145/1806799.1806811} {\path{doi:10.1145/1806799.1806811}}.

\bibitem{sun_towards_2011}
C.~Sun, D.~Lo, S.-C. Khoo, J.~Jiang, Towards more accurate retrieval of duplicate bug reports, in: 2011 26th {IEEE}/{ACM} International Conference on Automated Software Engineering ({ASE} 2011), 2011, pp. 253--262.
\newblock \href {https://doi.org/10.1109/ASE.2011.6100061} {\path{doi:10.1109/ASE.2011.6100061}}.

\bibitem{alipour_contextual_2013}
A.~Alipour, A.~Hindle, E.~Stroulia, A contextual approach towards more accurate duplicate bug report detection, in: 2013 10th Working Conference on Mining Software Repositories ({MSR}), 2013, pp. 183--192.
\newblock \href {https://doi.org/10.1109/MSR.2013.6624026} {\path{doi:10.1109/MSR.2013.6624026}}.

\bibitem{lazar_improving_2014}
A.~Lazar, S.~Ritchey, B.~Sharif, Improving the accuracy of duplicate bug report detection using textual similarity measures, in: Proceedings of the 11th Working Conference on Mining Software Repositories, {MSR} 2014, 2014, pp. 308--311.
\newblock \href {https://doi.org/10.1145/2597073.2597088} {\path{doi:10.1145/2597073.2597088}}.

\bibitem{ebrahimi_hmm-based_2019}
N.~Ebrahimi, A.~Trabelsi, M.~S. Islam, A.~Hamou-Lhadj, K.~Khanmohammadi, An {HMM}-based approach for automatic detection and classification of duplicate bug reports 113  98--109.
\newblock \href {https://doi.org/10.1016/j.infsof.2019.05.007} {\path{doi:10.1016/j.infsof.2019.05.007}}.

\bibitem{liu_fill_2023}
Z.~Liu, C.~Chen, J.~Wang, X.~Che, Y.~Huang, J.~Hu, Q.~Wang, Fill in the {Blank}: {Context}-aware {Automated} {Text} {Input} {Generation} for {Mobile} {GUI} {Testing}, in: 2023 {IEEE}/{ACM} 45th {International} {Conference} on {Software} {Engineering} ({ICSE}), 2023, pp. 1355--1367.
\newblock \href {https://doi.org/10.1109/ICSE48619.2023.00119} {\path{doi:10.1109/ICSE48619.2023.00119}}.

\bibitem{kang_large_2023}
S.~Kang, J.~Yoon, S.~Yoo, Large {Language} {Models} are {Few}-shot {Testers}: {Exploring} {LLM}-based {General} {Bug} {Reproduction}, in: 2023 {IEEE}/{ACM} 45th {International} {Conference} on {Software} {Engineering} ({ICSE}), 2023, pp. 2312--2323, iSSN: 1558-1225.
\newblock \href {https://doi.org/10.1109/ICSE48619.2023.00194} {\path{doi:10.1109/ICSE48619.2023.00194}}.

\bibitem{liu_chatting_2023}
Z.~Liu, C.~Chen, J.~Wang, M.~Chen, B.~Wu, X.~Che, D.~Wang, Q.~Wang, Chatting with {GPT}-3 for {Zero}-{Shot} {Human}-{Like} {Mobile} {Automated} {GUI} {Testing} (2023).
\newblock \href {https://doi.org/10.48550/arXiv.2305.09434} {\path{doi:10.48550/arXiv.2305.09434}}.

\bibitem{feldt_towards_2023}
R.~Feldt, S.~Kang, J.~Yoon, S.~Yoo, Towards {Autonomous} {Testing} {Agents} via {Conversational} {Large} {Language} {Models}, in: 2023 38th {IEEE}/{ACM} {International} {Conference} on {Automated} {Software} {Engineering} ({ASE}), 2023, pp. 1688--1693.
\newblock \href {https://doi.org/10.1109/ASE56229.2023.00148} {\path{doi:10.1109/ASE56229.2023.00148}}.

\bibitem{yu_llm_2023}
S.~Yu, C.~Fang, Y.~Ling, C.~Wu, Z.~Chen, Llm for test script generation and migration: Challenges, capabilities, and opportunities, in: 2023 IEEE 23rd International Conference on Software Quality, Reliability, and Security (QRS), 2023, pp. 206--217.
\newblock \href {https://doi.org/10.1109/QRS60937.2023.00029} {\path{doi:10.1109/QRS60937.2023.00029}}.

\bibitem{macneil_generating_2022}
S.~MacNeil, A.~Tran, D.~Mogil, S.~Bernstein, E.~Ross, Z.~Huang, Generating {Diverse} {Code} {Explanations} using the {GPT}-3 {Large} {Language} {Model}, in: Proceedings of the 2022 {ACM} {Conference} on {International} {Computing} {Education} {Research} - {Volume} 2, Vol.~2 of {ICER} '22, New York, NY, USA, 2022, pp. 37--39.
\newblock \href {https://doi.org/10.1145/3501709.3544280} {\path{doi:10.1145/3501709.3544280}}.

\bibitem{siddiq_zero-shot_2023}
M.~L. Siddiq, A.~Samee, S.~R. Azgor, M.~A. Haider, S.~I. Sawraz, J.~C.~S. Santos, Zero-shot {Prompting} for {Code} {Complexity} {Prediction} {Using} {GitHub} {Copilot}, in: 2023 {IEEE}/{ACM} 2nd {International} {Workshop} on {Natural} {Language}-{Based} {Software} {Engineering} ({NLBSE}), 2023, pp. 56--59.
\newblock \href {https://doi.org/10.1109/NLBSE59153.2023.00018} {\path{doi:10.1109/NLBSE59153.2023.00018}}.

\bibitem{mashhadi_method-level_2023}
E.~Mashhadi, H.~Ahmadvand, H.~Hemmati, Method-{Level} {Bug} {Severity} {Prediction} using {Source} {Code} {Metrics} and {LLMs}, in: 2023 {IEEE} 34th {International} {Symposium} on {Software} {Reliability} {Engineering} ({ISSRE}), 2023, pp. 635--646.
\newblock \href {https://doi.org/10.1109/ISSRE59848.2023.00055} {\path{doi:10.1109/ISSRE59848.2023.00055}}.

\bibitem{purba_software_2023}
M.~D. Purba, A.~Ghosh, B.~J. Radford, B.~Chu, Software {Vulnerability} {Detection} using {Large} {Language} {Models}, in: 2023 {IEEE} 34th {International} {Symposium} on {Software} {Reliability} {Engineering} {Workshops} ({ISSREW}), 2023, pp. 112--119.
\newblock \href {https://doi.org/10.1109/ISSREW60843.2023.00058} {\path{doi:10.1109/ISSREW60843.2023.00058}}.

\bibitem{weng_automatic_2023}
G.~Weng, A.~Andrzejak, Automatic {Bug} {Fixing} via {Deliberate} {Problem} {Solving} with {Large} {Language} {Models}, in: 2023 {IEEE} 34th {International} {Symposium} on {Software} {Reliability} {Engineering} {Workshops} ({ISSREW}), 2023, pp. 34--36.
\newblock \href {https://doi.org/10.1109/ISSREW60843.2023.00040} {\path{doi:10.1109/ISSREW60843.2023.00040}}.

\bibitem{petrovic_chatgpt_2023}
N.~Petrović, S.~Koničanin, S.~Suljović, {ChatGPT} in {IoT} {Systems}: {Arduino} {Case} {Studies}, in: 2023 {IEEE} 33rd {International} {Conference} on {Microelectronics} ({MIEL}), 2023, pp. 1--4.
\newblock \href {https://doi.org/10.1109/MIEL58498.2023.10315791} {\path{doi:10.1109/MIEL58498.2023.10315791}}.

\bibitem{ding_hpc-gpt_2023}
X.~Ding, L.~Chen, M.~Emani, C.~Liao, P.-H. Lin, T.~Vanderbruggen, Z.~Xie, A.~Cerpa, W.~Du, {HPC}-{GPT}: {Integrating} {Large} {Language} {Model} for {High}-{Performance} {Computing}, in: Proceedings of the {SC} '23 {Workshops} of {The} {International} {Conference} on {High} {Performance} {Computing}, {Network}, {Storage}, and {Analysis}, {SC}-{W} '23, 2023, pp. 951--960.
\newblock \href {https://doi.org/10.1145/3624062.3624172} {\path{doi:10.1145/3624062.3624172}}.

\bibitem{koreeda_larch_2023}
Y.~Koreeda, T.~Morishita, O.~Imaichi, Y.~Sogawa, {LARCH}: {Large} {Language} {Model}-based {Automatic} {Readme} {Creation} with {Heuristics}, in: Proceedings of the 32nd {ACM} {International} {Conference} on {Information} and {Knowledge} {Management}, {CIKM} '23, 2023, pp. 5066--5070.
\newblock \href {https://doi.org/10.1145/3583780.3614744} {\path{doi:10.1145/3583780.3614744}}.

\end{thebibliography}

\end{document}